\documentclass{jfm}

\usepackage{graphicx}
\usepackage{natbib}
\usepackage{amsmath}

\ifCUPmtlplainloaded \else
  \checkfont{eurm10}
  \iffontfound
    \IfFileExists{upmath.sty}
      {\typeout{^^JFound AMS Euler Roman fonts on the system,
                   using the 'upmath' package.^^J}%
       \usepackage{upmath}}
      {\typeout{^^JFound AMS Euler Roman fonts on the system, but you
                   dont seem to have the}%
       \typeout{'upmath' package installed. JFM.cls can take advantage
                 of these fonts,^^Jif you use 'upmath' package.^^J}%
      }
  \else
  \fi
\fi

\ifCUPmtlplainloaded \else
  \checkfont{msam10}
  \iffontfound
    \IfFileExists{amssymb.sty}
      {\typeout{^^JFound AMS Symbol fonts on the system, using the
                'amssymb' package.^^J}%
       \usepackage{amssymb}%
       \let\le=\leqslant  
       \let\ge=\geqslant  \let\geq=\geqslant
      }{}
  \fi
\fi

\ifCUPmtlplainloaded \else
  \IfFileExists{amsbsy.sty}
    {\typeout{^^JFound the 'amsbsy' package on the system, using it.^^J}%
     \usepackage{amsbsy}}
    {\providecommand\boldsymbol[1]{\mbox{\boldmath $##1$}}}
\fi

\newcommand\Rey{\mbox{\textit{Re}}}  

\newsavebox{\astrutbox}
\sbox{\astrutbox}{\rule[-5pt]{0pt}{20pt}}

\newcommand\be{\boldsymbol{e}}

\newcommand\hbu{\mathbf{\hat{u}}}
\newcommand\bu{\mathbf{u}}
\newcommand\bone{\boldsymbol{1}}
\newcommand\bsigma{\boldsymbol{\sigma}}
\newcommand\hbsigma{\boldsymbol{\hat{\sigma}}}
\newcommand\hp{\hat{p}}
\newcommand\bn{\boldsymbol{n}}
\newcommand\bfn{\boldsymbol{\mathrm{f}}_n}

\title[Inertial migration of a rigid sphere in three-dimensional Poiseuille flow]{Inertial migration of a rigid sphere in three-dimensional Poiseuille flow}

\author[K. Hood, S. Lee and M. Roper]%
{Kaitlyn Hood$^1$%
  \thanks{Email address for correspondence: kaitlyn.t.hood@gmail.com},\ns
Sungyon Lee$^2$,
and Marcus Roper$^1$}

 \affiliation{$^1$Department of Mathematics, University of California,
 Los Angeles, CA 90095, USA\\[\affilskip]
$^2$Department of Mechanical Engineering, Texas A\&M University,
 College Station, TX 77843, USA}

\pubyear{}
\volume{}
\pagerange{}
\date{?; revised ?; accepted ?. - To be entered by editorial office}
\begin{document}

\maketitle

\begin{abstract}
Inertial lift forces are exploited within inertial microfluidic devices to position, segregate, and sort particles or droplets. However the forces and their focusing positions can currently only be predicted by numerical simulations, making rational device design very difficult. Here we develop theory for the forces on particles in microchannel geometries. We use numerical experiments to dissect the dominant balances within the Navier-Stokes equations and derive an asymptotic model to predict the lateral force on the particle as a function of particle size.  Our asymptotic model is valid for a wide array of particle sizes and Reynolds numbers, and allows us to predict how focusing position depends on particle size.
\end{abstract}


\section{Introduction}\label{sec:intro}

Inertial microfluidic devices employ inertial focusing to segregate and sort chains of particles, and to move particles between streams of different fluids.  For example, centrifuges-on-a-chip \citep{DiCarlo11, DiCarlo14} trap circulating cancer cells from blood in microchannel vortices, and sheathless high-throughput flow cytometry \citep{DiCarlo10, DiCarlo13} fractionates particles from a buffer in order to image and count rare blood cells. However there are no predictive theories that describe the trajectories of particles during inertial focusing. Instead the features of these devices, including flow-rate and geometry, are optimized by experimental trial-and-error. Although asymptotic theories exist for inertial lift forces, they are quantitatively correct only for asymptotically small particles, much smaller than the particles that are typically used in microfluidic devices. Previous asymptotic theories also do not predict how differently sized particles will be differently focused \citep{DiCarlo09}.

Inertial migration of particles was first observed in 1961 by Segr\'{e} and Silberberg. Experiments showed that a dilute suspension of neutrally buoyant particles flowing in a cylindrical pipe at moderate speeds will migrate across streamlines \citep{SegreSilberberg61, SegreSilberberg62a, SegreSilberberg62b}.  Particles initially uniformly dispersed through the cross-section of the pipe became focused into a ring with radius 0.6 times the channel radius.  Since the reversibility of Stokes equations (the limit of the Navier-Stokes equations when Reynolds number, $\Rey=0$) prohibits movement across streamlines this migration must arise from inertia in the flow \citep{Bretherton62}.  

Many theoretical studies of this effect using asymptotic theory are described below.  Each study focuses on a particular limit of two dimensionless groups, $\Rey$ and $\alpha$.  The first parameter, $\Rey$, is the channel Reynolds number, and only depends on the dimensions of the pipe and the properties of unladen flow into the channel.  The second parameter, $\alpha$, is a ratio of the particle size to a characteristic channel length scale. Some studies take this length scale to be the width of the channel, others the distance between the particle and the wall. Values for these parameters in various studies are compiled in table \ref{tab:prevresults}.

Although early theoretical studies \citep{RubinowKeller61,Saffman65} illuminated how inertial lift forces are generated by applied torques or body forces, \cite{CoxBrenner68} were the first to directly address lift forces on neutrally buoyant particles. They consider a body of arbitrary shape suspended in a fluid bounded by a system of walls in three dimensions, and observe that viscous stresses dominate over inertial stresses, provided that $\Rey \ll \alpha$.  Assuming rapid flow field decay, i.e. viscous stresses remain dominant over inertial stresses throughout the fluid, they derive an implicit analytic expression for the force by a regular perturbation expansion of the Navier-Stokes equations in the small parameter $\Rey$.  They show that this assumption is valid for the lateral migration of a sphere in flow through a cylinder with arbitrary cross section. Subsequently, they arrive at an integral formula for the lift force for a neutrally buoyant sphere, but they do not evaluate the integrals to determine how lift forces vary across the channel, or how they depend on particle size.  Additionally, \cite{McLaughlin91} extended the theory of \cite{Saffman65} to non-neutrally buoyant particles by considering a finite slip velocity.  

\cite{HoLeal74} were the first to explicitly calculate the lift force on a particle in the presence of channel walls, by developing an asymptotic theory for a particle in 2D Couette and Poiseuille flows.  Since there are multiple scales for the dynamics in the particle-channel system, Ho \& Leal introduce the particle Reynolds number $\Rey_p = \alpha^2 \Rey$.  They observe that provided $\Rey_p \ll \alpha^2$, viscous stresses dominate over inertial stresses throughout the fluid filled domain.  They develop a scaling law for the lift force as a function of the particle position by a regular perturbation series expansion in powers of $\Rey_p$.   Each term in this expansion can be expanded in powers of $\alpha$. Retaining only leading order terms, they find that lift force $F_L \sim \rho U_m^2\alpha^2 a^2$, where $\rho$ is the fluid density, $U_m$ is the maximum velocity of the background flow and $a$ is the particle radius, i.e. that lift force scales with the fourth power of particle diameter. 

Later computations by \cite{VasseurCox76} apply the result of \cite{CoxBrenner68} to a spherical particle flowing between two parallel plates.  Provided $\Rey_p \ll \alpha^2$, only the inner expansion is needed to calculate the first term in the expansion for the migration velocity.  The migration velocity is computed as a Fourier integral and no definite scaling law for the lift force is derived. However, they compare their numerical results to those of \cite{HoLeal74} and have good agreement, except near the wall. Similarly, by considering a particle near a single wall and using the results of \cite{CoxBrenner68}, \cite{CoxHsu77} calculate the migration velocity of a particle near a the wall.  They do not derive a scaling law for the force, but their numerical results compare well to those of \cite{HoLeal74} near the wall.

\begin{table}
\centering 
\begin{tabular}{llllll}
study & $\alpha$ & $\Rey$ & $\Rey_p$ & $p$ & Comments \\ \hline
\cite{RubinowKeller61} & N/A & N/A & $\ll1$ & 5 & Uniform flow and absence of walls \\
\cite{Saffman65} & N/A & N/A & $\ll1$ & 2 & Wall effect: particle lags behind fluid\\
\cite{CoxBrenner68} & $\ll1$ & $\ll1$ & $\ll\alpha^2$ & - & Implicit analytic force expression \\
\cite{HoLeal74} & $\ll1$ & $\ll1$ & $\ll\alpha^2$ & 4 & 2D geometry \\
\cite{VasseurCox76} & $\ll1$ & $\ll1 $ & $\ll\alpha^2$ & - & Agrees with Ho \& Leal away from wall \\
\cite{CoxHsu77} & $\ll1$ & $\ll1$ & $\ll\alpha^2$ & - & Agrees with Ho \& Leal near wall \\
\cite{Hinch89} & $\ll1$ & $O(1)$ & $\ll1$ & 4 & Matched asymptotics   \\
\cite{McLaughlin91} & N/A & N/A & $\ll1$ & 2 & Extends Saffman for finite slip velocity \\
\cite{Hogg94} & $\ll1$ & $O(1)$ & $\ll\alpha$ & 4 & Studies non-neutrally buoyant particles \\
\cite{Asmolov99} & $\ll1$ & $O(10^3)$ & $\ll1$ & 4 & Extends Schonberg \& Hinch for large $\Rey$ \\
\cite{DiCarlo09} & $O(1)$ & $O(10^2)$ & $O(10)$ & 3 & 3D numerics and experiments \\
This paper & $O(1)$ & $O(10^2)$ & $O(10)$ & - & Reconciles with $\alpha \ll 1$ theory \\
\hline
\end{tabular}
\caption{A comparison of the parameters $\alpha$, $\Rey$, and $\Rey_p$, and the value of the exponent $p$ for the scaling law $f \sim \rho U^2 a^p$, for various studies, where $\rho$ is the fluid density, $U$ is the characteristic flow velocity, and $a$ is the particle radius.}
\label{tab:prevresults}
\end{table}

Although early theory assumed $\Rey\ll1$, in inertial microfluidic devices, and in the experiments of \cite{SegreSilberberg61}, the channel Reynolds number ranges from 1-700. The first theory capable of describing migration of particles in these moderate Reynolds number flows was developed by \cite{Hinch89} who assumed small particle size ($\alpha \ll 1$) and particle Reynolds number ($\Rey_p = \alpha^2 \Rey \ll 1$), but allowed for Reynolds number $\Rey = O(1)$. For particles in a 2D Poiseuille flow, they separate the flow field into inner and outer regions. In the inner region, at distances $O(a)$ from the particle, the viscous stresses are dominant.  In the outer region, at distances $a/\Rey_p^{-1/2}$ from the particle, inertial stresses become co-dominant with viscous stresses. In this outer region, the particle's disturbance of the flow field is weak enough to be linearized around the base flow, reducing the Navier-Stokes equation to Oseen's linearized equations \citep{Batchelor67}. Although the authors solve for the inertial migration velocity for a force free particle, their calculation can readily be adapted to calculate the lift force, and again predicts $f_L \sim \rho U^2 \alpha^2 a^2$; i.e. that lift force scales with the fourth power of particle size. \cite{Hogg94} extended the analysis of \cite{Hinch89} to non-neutrally buoyant particles, while \cite{Asmolov99} extended the theory of \cite{Hinch89} to large $\Rey$.

In inertial microfluidic experiments particle diameters may not be small compared to the channel width and particle Reynolds numbers $\Rey_p$ can reach values of 10-20. To determine lift forces in this experimentally relevant regime, and to consider focusing in three-dimensional flows, \cite{DiCarlo09} performed finite element simulations for particles in square channels.  They varied Reynolds $\Rey$ number between 20 and 80 and the ratio of particle size to channel size $\alpha$ between 0.05 and 0.2.  They find that unlike circular pipes, which focus particles to an annulus, square channels focus particles to four symmetrically arranged positions. For particles near the channel center, numerical fitting of the numerical data generates the power law $f_L \sim \rho U^2 \alpha a^2$, asserting that the lift force $f_L$ increases with $a^3$ rather than $a^4$. For particles closer to the channel walls they find different exponents for the scaling of lift force with particle size, depending on particle position.  The different exponent in the scaling casts doubt on the use of any of the previous asymptotic theories. Additionally, \cite{DiCarlo09} explore experimentally and numerically how the focusing position of the particle varies with particle size; an observation that is integral to inertial separation devices, but which is not considered in asymptotic theory.

In this paper we explicitly compute the dominant balances in the equations of motion of the particle to show that the asymptotics of \cite{HoLeal74} were essentially correct, and hold for a much larger parameter space of $\Rey$ and $\alpha$ than the authors realized.  Specifically, viscous and pressure stresses dominate over inertial stresses over the entire width of the channel; and the drag force on the particle can be computed by regular perturbation of the equations of slow creeping flow. We perform this regular perturbation analysis to derive asymptotic expressions for the lift force that are quantitatively accurate up to $\Rey = 80$, and with maximum particle size limited only by the proximity of the walls. Our theory also predicts how focusing position depends on particle radius. We show that the scaling observed by \cite{DiCarlo09} is actually a serendipitous fitting to a perturbation series in $\alpha$ by a single apparent scaling law.

We organize the paper as follows; in \S \ref{sec:eqns} we formulate and solve numerically for the inertial lift force on a drag-free spherical particle, focusing on the dependence of this lift force on particle size and channel Reynolds number. In \S \ref{sec:oseen} we dissect out the dominant balances in these equations. In \S \ref{sec:asymptotics} we develop a regular perturbation series for the lift force, similar to that of \citet{HoLeal74}, and show that it is in quantitatively good agreement with the numerically computed lift force (Fig. \ref{fig:channel}b).  In \S \ref{sec:3d} we describe how we generalize the computation to three dimensional channel flows, and in \S \ref{sec:discussion} we show good agreement of our asymptotic method with experiments and discuss its possible applications.

\section{Equations of motion}\label{sec:eqns}

\begin{figure}
  \centerline{
       $(a)$\includegraphics{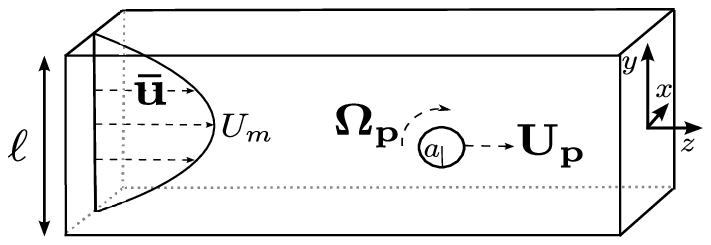}
       $(b)$\includegraphics{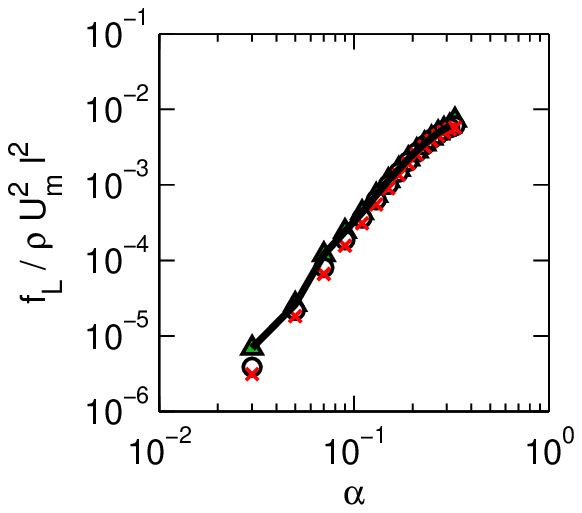}
    }
  \caption{$(a)$ The physical system for the flow around a particle suspended in a square channel. $(b)$ We numerically compute the lift force $f_L$ as a function of particle size $\alpha$ for various Reynolds numbers, $\Rey=10$ (green triangles), $\Rey=50$ (circles), and $\Rey=80$ (red x's).  The curves collapse when lift force is scaled by $\rho U_m^2 \ell^2$, but the curves are neither a power law with exponent 3 nor exponent 4.  A regular perturbation expansion that we computed numerically fits the data extremely well (solid black line).}
\label{fig:channel}
\end{figure}

We model flow through an infinitely long square channel of side length $\ell$.  A three dimensional Poiseuille flow $\bar{\mathbf{u}}'$ flowing in the $z'-$direction, is disturbed by a rigid sphere of radius $a$ (figure \ref{fig:channel}$a$). Here we use primes to denote dimensional variables.  We denote the fluid viscosity by $\mu$, fluid density by $\rho$, and the center-line velocity of the background flow by $U_m$.  The particle is located at $(x_0',y_0',0)$ and is allowed to translate in the $z'-$ direction with velocity $\mathbf{U}_p'=U_p'\mathbf{e_{z'}}$, and rotate with angular velocity $\mathbf{\Omega}_p'$, until it is drag free and torque free.  The objective of this paper is to calculate the lift forces acting on the particle in the $x'-$ and $y'-$ directions.

There are three important dimensionless parameters: (i) the dimensionless ratio of particle radius to channel diameter $\alpha=a/\ell$, (ii) the channel Reynolds number $\Rey= U_m\ell/\nu$, and (iii) the particle Reynolds number $\Rey_p =  U_m a^2/\ell \nu$. Here we write $\nu = \mu/\rho$ for the kinematic viscosity. In common with previous theory \citep{CoxBrenner68, HoLeal74, Hinch89} we will perform dual perturbation expansions in $\Rey_p$ and $\alpha$, assuming that both quantities are asymptotically small. In inertial microfluidic experiments \citep{DiCarlo09}, particle diameters may be comparable with the channel dimensions. We will show that our expansions converge even at the moderate values of $\alpha$ accessed in these experiments.  

The background flow, $\mathbf{\bar{u}}'$, is square channel Poiseuille flow \citep{ Papanastasiou99}, and takes the form $\mathbf{\bar{u}}'=  \bar{u}'(x',y')\mathbf{e}_{z'}$, where $\bar{u}'$ defined by:
\begin{align}\label{eq:ubar}
  \bar{u}'(x',y') = U_m \Bigg[  &-\frac{1}{2} \left(y'^2 - \left(\frac{\ell}{2a}\right)^2\right) \\
  &+ \sum_{n=0}^\infty  \frac{-4 \ell^2 (-1)^n \cosh\left(\frac{(2n+1)\pi a x'}{\ell}\right)}{(2n+1)^3\pi^3 a^2 \cosh\left(\frac{(2n+1)\pi }{2}\right)} \cos\left(\frac{(2n+1)\pi a y'}{\ell}\right) \Bigg]. \nonumber
\end{align}
The velocity $\mathbf{\bar{u}}'$ and pressure $\bar{p}'$ solve the Stokes equations with boundary condition $\mathbf{\bar{u}}' = \mathbf{0}$ on the channel walls.  We will also need the Taylor series expansion for $\bar{u}'$ around the center of the particle:
\begin{align}\label{eq:ubarseries}
  \bar{u}'(x',y') = \beta' &+ \gamma_x' (x'-x_0') + \gamma_y' (y'-y_0')  \\ \nonumber &+ \delta_{xx}' (x'-x_0')^2 + \delta_{xy}' (x'-x_0')(y'-y_0') +\delta_{yy}' (y'-y_0')^2 + O(r'^3) 
\end{align}

To illustrate the reference frame of the equations we will use later, we first list the dimensionless equations of motion and boundary conditions for the velocity and pressure fields $\mathbf{u}''$ and $p''$ expressed in particle-fixed coordinates.  We non-dimensionalise these equations by scaling velocities by $U_m a/\ell$, lengths by $a$, and pressures by $\mu U_m / \ell$:

\begin{eqnarray}
  \nabla^2\mathbf{u}'' - \nabla p'' &=& \Rey_p \, [(\mathbf{u}''+\mathbf{U}_p)\cdot\nabla \mathbf{u}''], \nonumber \\
  \nabla \cdot \mathbf{u}'' &=& 0, \nonumber \\
  \mathbf{u}'' &=& \mathbf{\Omega}_p \times \mathbf{r''} \quad \mathrm{on} \quad r''=1, \\
  \mathbf{u}'' &=& -\mathbf{U}_p \quad \mbox{on the walls}, \nonumber \\
  \mathbf{u}'' &=& \mathbf{\bar{u}} - \mathbf{U}_p \quad \mathrm{as} \quad z'' \to \pm \infty.\nonumber
\end{eqnarray}

Now we introduce the disturbance velocity and pressure fields $\mathbf{u} = \mathbf{u}''-\mathbf{\bar{u}} + \mathbf{U}_p$ and $p = p'' - \bar{p}$, in which the background flow $\mathbf{\bar{u}} - \mathbf{U}_p$ (as measured in this reference frame) is subtracted from $\mathbf{u}''$. For reference, the fluid velocity in the lab frame is given by: $\mathbf{v}=\mathbf{u}+\mathbf{\bar{u}}$. We then obtain the equations of motion and boundary conditions that will be used throughout this paper:

\begin{eqnarray} \label{eq:NSE_diff}
  \nabla^2\mathbf{u} - \nabla p &=& \Rey_p \, (\mathbf{\bar{u}}\cdot\nabla \mathbf{u} + \mathbf{u}\cdot\nabla \mathbf{\bar{u}} + \mathbf{u}\cdot \nabla\mathbf{u}),  \nonumber \\
  \nabla \cdot \mathbf{u} &=& 0, \nonumber \\
  \mathbf{u} &=& \mathbf{\Omega_p} \times \mathbf{r} - \mathbf{\bar{u}} + \mathbf{U_p} \quad \mathrm{on} \quad r=1, \\
  \mathbf{u} &=& \mathbf{0} \quad \mbox{on the walls}, \nonumber \\
  \mathbf{u} &=& \mathbf{0} \quad \mathrm{as} \quad z \to \pm \infty. \nonumber
\end{eqnarray}

We call the variables that appear in (\ref{eq:NSE_diff}) the inner variables. Appendix \ref{appNotation} summarizes the notations used for dimensionless and dimensional variables. 

We formulated (\ref{eq:NSE_diff}) as a finite element model (FEM) with $\sim 650,000$ linear tetrahedral elements, and solved for $\mathbf{u}$ and $p$ using Comsol Multiphysics (COMSOL, Los Angeles) in a rectangular domain with dimensions $\frac{\ell}{a} \times \frac{\ell}{a} \times 5 \frac{\ell}{a}$, prescribing $\mathbf{u}$ at the inlet $z=-5\frac{\ell}{a}$, and imposing neutral boundary conditions (vanishing stress) at the outlet $z=+5\frac{\ell}{a}$. In the FEM, we vary $U_p$ and $\mathbf{\Omega}_p$ until there is no drag force or torque on the particle.  The FEM Lagrange multipliers, which enforce the velocity boundary condition on the particle, are used to compute the lift force $f_L$ on the drag free and force free particle. \citet{bramble81} rigorously demonstrates the accuracy of flux calculations from Lagrange multipliers for a Poisson's equation with Dirichlet boundary conditions. Additionally, we discuss accuracy tests of the FEM discretization for our problem in Appendix \ref{appModel}. 

First, we consider the lift force for particles located on the line of symmetry $x_0 = 0$. Fixing particle position $y_0$, we found curves of lift force $f_L$ against particle size $a$ collapsed for different Reynolds numbers.  Particles in different positions have different apparent scaling's for $f_L$ as a function of $a$ (figure \ref{fig:scalinglaw}$a$-$b$). By assaying a large range of particle sizes $\alpha$ we see that the empirical fit $f_L\sim \rho U^2a^3$ observed by \citet{DiCarlo09} is not asymptotic as $a\to 0$. The data for smallest particle sizes ($\alpha<0.07$) are consistent with a scaling law of $f_L\sim \rho U^2 a^4/\ell^2$ as predicted by \citet{HoLeal74} and \cite{Hinch89}, but extrapolation of the asymptotic force law to the moderate particle sizes used in real inertial microfluidic devices ($\alpha\approx 0.1-0.3$) over-predicts the lift force by more than an order of magnitude.

\section{Dominant balances in the equations of motion}\label{sec:oseen}

The governing equation (\ref{eq:NSE_diff}) is a balance between momentum flux and the pressure and viscous stresses. Testing the hypothesis that two of these three contributions might form a dominant balance within the equation, we plotted the resultants of the three fluxes as functions of distance from the particle. Specifically, we integrate the $\ell_2$ norm of each flux over spherical control surfaces centered at the particle.  Let $S_r$ be the boundary of a sphere of radius $r$ centered at the origin, and define the $\ell_2$ norm by $\| \mathbf{u} \|_2 = \sqrt{u^2 + v^2 +w^2}$. Then the dimensionless viscous stress resultant acting on the sphere $S_r$ is defined by:
\begin{equation}\label{eq:V}
  V(r) = \int_{S_r} \| \nabla \mathbf{u} \cdot \mathbf{n} \|_2 \mathrm{ds} ~,
\end{equation}
\noindent and the dimensionless inertial term $I(r)$ stress resultant by:
\begin{equation}\label{eq:I}
  I(r) = \Rey_p \int_{S_r} \| [ (\mathbf{\bar{u}}-\mathbf{U_p})\mathbf{u} + \mathbf{u} (\mathbf{\bar{u}}-\mathbf{U_p}) + \mathbf{u}\mathbf{u}]\cdot \mathbf{n} \|_2 \mathrm{ds}~
\end{equation}
The integrand in $I(r)$ is chosen to have divergence equal to the right hand side of (\ref{eq:NSE_diff}), and we pick a form of the inertial flux that decays in $\ell_2$ norm as $r\to \infty$.

\begin{figure}
  \begin{center}
    $(a)$\includegraphics{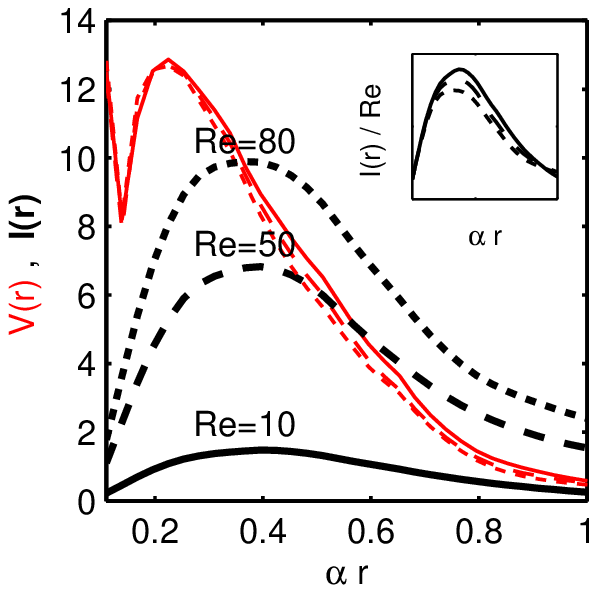}
    $(b)$\includegraphics{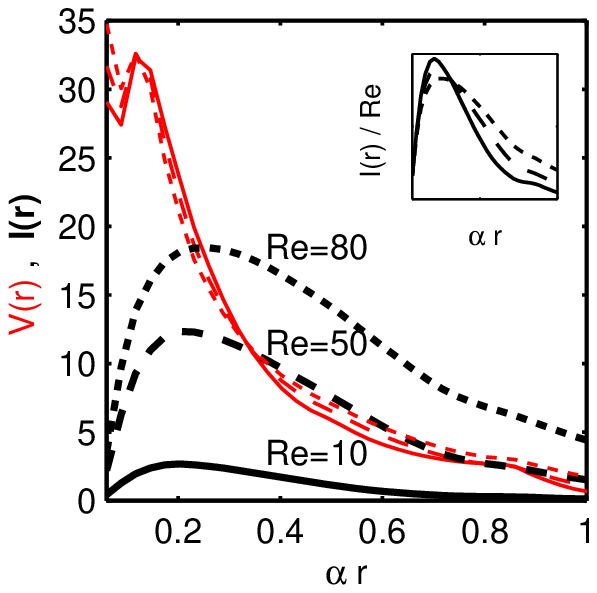}
    \caption{  We examine the dominant balance of the Navier-Stokes equation for $(a)$ a particle near the channel center ($y_0 = 0.15 / \alpha$ and $\alpha = 0.11$), and $(b)$ a particle near the channel walls ($y_0 = 0.35 / \alpha$ and $\alpha = 0.06$).  The viscous stresses $V(r)$ for various Reynolds numbers are plotted as thin red lines, and the inertial stresses $I(r)$ are plotted as thick black lines.  Reynolds numbers are indicated by line style, $\Rey=10$ solid line, $\Rey=50$ dashed line, and $\Rey=80$ dotted line.  The inset figures show that the inertial stresses $I(r)$ collapse when scaled by $\Rey$, suggesting that the high Reynolds number dynamics are determined by the low Reynolds number dynamics. }
   \label{fig:NSEterms}
  \end{center}
\end{figure}

Numerically evaluating these two terms as well as $\frac{1}{r} \int_{S_r} \| p \mathbf{n} \|_2\,{\rm d}s$ we find that contrary to the predictions of \cite{HoLeal74} and \cite{Hinch89} that at moderate channel Reynolds numbers, the viscous and pressure stress resultants are numerically larger than the momentum flux. In particular there is no region in which $V(r)$ and $I(r)$ are co-dominant at $\Rey=10$ (figure \ref{fig:NSEterms}). Indeed, even at higher Reynolds numbers ($\Rey=50$, 80) for which inertial stresses are numerically larger than viscous stresses, inertial stresses can be collapsed onto a single curve (see inset of figures \ref{fig:NSEterms}a-b) by rescaling with $\Rey$.  This scaling suggests that the underlying dynamics, even at moderate values of $\Rey$, are inherited from the small $\Rey$ dominant balance of pressure and viscous stresses. Dominance of viscous stresses over inertial stresses is surprising because as \cite{HoLeal74} noticed, the resulting dominant balance equations are not self-consistent for isolated particles in unbounded fluid flow.

We will now present a first order estimate of the size of the domain in which inertial stresses may be expected to be dominant. The slowest decaying component of the disturbance flow associated with a force free particle on the plane of symmetry ($x_0=0$) is given by the stresslet flow \citep{ Batchelor67, KimKarrila2005}:
\begin{equation}
\mathbf{u}_{\mathrm{stresslet}} = \frac{5 \gamma_y (y- y_0) z \mathbf{r}}{2 r^5} = O\left(\frac{1}{r^2}\right).
\end{equation}
Recall that $\gamma_y$ is the strain rate, defined in (\ref{eq:ubarseries}). For this flow field, the viscous stress term in (\ref{eq:NSE_diff}) decays with distance like:
\begin{equation}
V(r) \sim O(\nabla \mathbf{u}_{\mathrm{stresslet}}) \sim O\left(\frac{1}{r^3}\right)~,
\end{equation}
whereas the inertial stresses vary with distance like:
\begin{equation}
  I(r) \sim  O(\Rey_p(\mathbf{u}_{\mathrm{stresslet}})(\mathbf{\bar{u}}-\mathbf{U_p})) \sim O\left(\frac{\Rey_p}{r}\right).
\end{equation}
We define the cross-over radius, $r_*$, to be the distance at which the viscous and inertial stresses are comparable,
\begin{equation}\label{eq:rstar}
  r_* = O\left(\frac{1}{\Rey_p^{1/2}}\right) .
\end{equation}
In order to compare the cross-over radius to the width of the channel we consider when $\alpha r_* =  O(1/\Rey^{1/2})$ is equal to one. To ensure that viscous stresses dominate over inertial stresses over the channel cross-section (i.e. $\alpha r_* \gg 1$) Ho \& Leal restrict to cases where $\Rey \ll 1$.  The asymptotic analysis of \cite{Hinch89} allows that $\Rey = O(1)$, but at the cost of needing to separately model and match the flows at $O(1)$ distances from the particle where viscous stresses are dominant, and at $O(1/\Rey^{1/2})$ distances where inertial and viscous stresses must both be included in the dominant balance.

However, the predicted cross over radius falls short of the numerical cross over radius (fig \ref{fig:NSEterms}, table \ref{tab:oseen}).  There are two explanations for the dominance of viscous stresses over inertial in these experimental geometries. First, the above estimates do not consider the coefficients in the stresslet; merely the order of magnitude of the terms.  Second, although the stresslet describes the flow disturbance for a force free particle in an unbounded fluid, the leading order flow is considerably altered by the presence of the channel walls. Below we demonstrate that both explanations contribute to the dominance of viscous stresses throughout the channel cross section, pushing the cross-over radius $r^*$ out beyond the channel walls (table \ref{tab:oseen}).

\begin{figure}
  \begin{center}
    $(a)$ \includegraphics{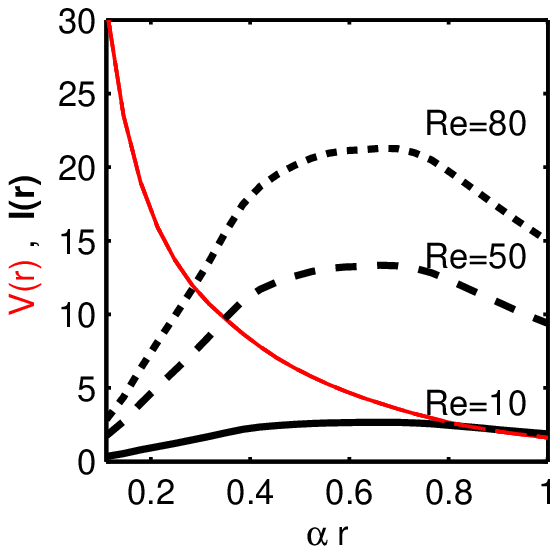}
    $(b)$ \includegraphics{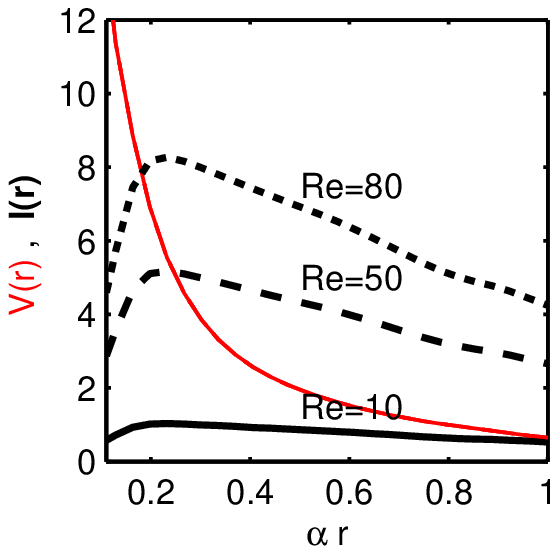}
    \caption{ We examine the dominant balance that arises from the stresslet approximation of the flow for $(a)$ a particle near the channel center ($y_0 = 0.15 / \alpha$ and $\alpha = 0.11$), and $(b)$ a particle near the channel walls ($y_0 = 0.35 / \alpha$ and $\alpha = 0.06$).  The viscous stresses $V(r)$ for various Reynolds numbers are plotted as thin red lines, and the inertial stresses $I(r)$ are plotted as thick black lines.  Reynolds numbers are indicated by line style, $\Rey=10$ solid line, $\Rey=50$ dashed line, and $\Rey=80$ dotted line. }
  \label{fig:NSEtermsStresslet}
 \end{center}
\end{figure}

\subsection{Role of the stresslet constants \label{sec:coeffs}}

We compute $I(r)$ and $V(r)$ numerically for the stresslet flow field (i.e. substitute $\mathbf{u} = \mathbf{u}_{\mathrm{stresslet}}$ in equations (\ref{eq:V}-\ref{eq:I})). We examine two representative cases; a medium sized particle near the channel center ($ y_0=0.15 / \alpha$, $\alpha = 0.11$) (figure \ref{fig:NSEtermsStresslet}$a$), and a small particle near the channel wall ($y=0.35 / \alpha$, $\alpha = 0.06$) (figure \ref{fig:NSEtermsStresslet}$b$). For $\Rey = 10$, in both cases the inertia is significantly smaller than the viscous stress throughout the channel. At larger values of $\Rey$, $I(r)$ eventually exceeds $V(r)$, but the cross-over radius $r_*$ is much larger than simple order of magnitude estimates would suggest (table \ref{tab:oseen}).

\subsection{Role of Wall Effects}  

\begin{figure}
  \begin{center}
    $(a)$ \includegraphics{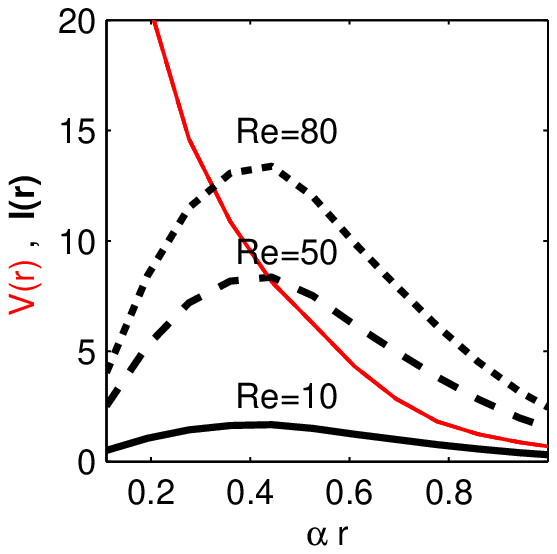}
    $(b)$ \includegraphics{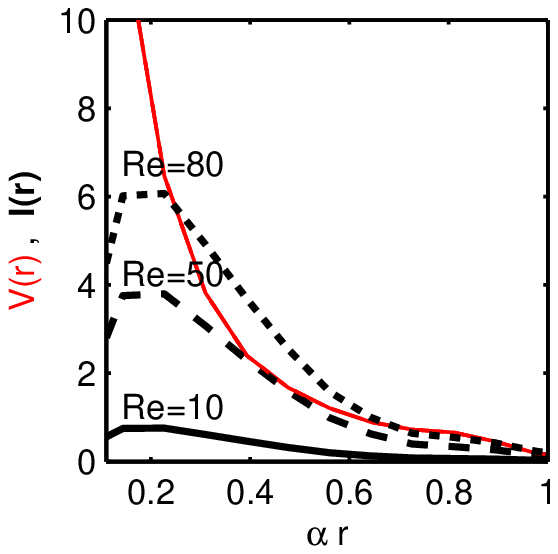}
    \caption{We examine the dominant balance that arises from the stresslet and first wall correction of the flow for $(a)$ a particle near the channel center ($ y_0 = 0.15 / \alpha$ and $\alpha = 0.11$), and $(b)$ a particle near the channel walls ($ y_0 = 0.35 / \alpha$ and $\alpha = 0.06$).  The viscous stresses $V(r)$ for various Reynolds numbers are plotted as thin red lines, and the inertial stresses $I(r)$ are plotted as thick black lines.  Reynolds numbers are indicated by line style, $\Rey=10$ solid line, $\Rey=50$ dashed line, and $\Rey=80$ dotted line.}
  \label{fig:NSEtermsWalleffect}
 \end{center}
\end{figure}

To estimate how wall modifications of the disturbance flow affect the dominant balances in Eq.(\ref{eq:NSE_diff}), we numerically computed the first wall correction.  That is, we substitute into equations (\ref{eq:V}-\ref{eq:I}) $\mathbf{u} = \mathbf{u}_{\mathrm {stresslet}} + \mathbf{u}_{\mathrm{image}}$, where $\mathbf{u}_{\mathrm{image}}$ is a solution of Stokes' equations with boundary condition $\mathbf{u}_{\mathrm{image}}=-\mathbf{u}_{\mathrm{stresslet}}$ on the channel walls.  We examine the same two representative cases as in \S \ref{sec:coeffs}: $(y_0=0.15 / \alpha, \alpha = 0.11)$ and $(y_0=0.35 / \alpha, \alpha = 0.06)$  (figure \ref{fig:NSEtermsWalleffect}$a-b$). For $\Rey = 10$, in both cases the inertia is significantly smaller than the viscous stress throughout the channel. At larger values of $\Rey$, $I(r)$ eventually exceeds $V(r)$, but the cross-over radius $r_*$ is larger than that predicted from the stresslet coefficients (table \ref{tab:oseen}).

We can rationalize the larger values of the cross-over radius $\alpha \, r_*$ by considering the boundary conditions on the channel walls.  Because the velocity field $\mathbf{u}$ vanishes on the channel walls, the inertial stresses vanish there. $I(r)$ is therefore suppressed at larger radii. We see less suppression of $V(r)$, presumably because viscous stresses do not need to vanish on the channel walls. Suppression of $I(r)$ increases the cross-over radius at which inertial stresses must be considered in the dominant balance.

\section{A series expansion for the inertial lift force}\label{sec:asymptotics}

Our careful evaluation of the stresslet prefactors and wall-contributions shows that viscous stresses are dominant over inertial stresses over much of the fluid filled domain, including at much greater distances from the particle than previous estimates have suggested. We therefore develop an asymptotic theory, based on \citet{CoxBrenner68} and \citet{HoLeal74}, in which the flow field, $\mathbf{u}$, pressure, $p$, particle velocity $\mathbf{U_p}$, and rotation $\mathbf{\Omega_p}$ are expanded in powers of $\Rey_p$, with inertia completely neglected in the leading order equations:
\begin{equation}
  \mathbf{u} = \mathbf{u}^{(0)} + \Rey_p \mathbf{u}^{(1)} + \ldots ~~, \quad
  p = p^{(0)} + \Rey_p p^{(1)} + \ldots ~~, \qquad \hbox{etc}. \\
\end{equation}

Notice that this is an expansion in the particle Reynolds number $\Rey_p$ and not the channel Reynolds number $\Rey$.  Although in experiments the channel Reynolds number is typically large, the expansion is formally valid provided that $\alpha^2$ is small enough that $\Rey_p = \alpha^2 \Rey \lesssim 1$. In fact when we compare our theory with numerical simulations in \S \ref{sec:results} we find that the perturbative series gives a good approximation to the lift force even for $\Rey_p = 7$ (Fig. \ref{fig:channel}b).

\begin{table}
\centering 
\begin{tabular}{l llllll} 
Cross over radius $\alpha r_* $ &\multicolumn{3}{c}{$y_0=0.15 / \alpha$} &\multicolumn{3}{c} {$y_0=0.35 / \alpha$}\\
  &\multicolumn{3}{c}{$\alpha=0.11$} &\multicolumn{3}{c} {$\alpha=0.06$}\\ [0.5ex] 
\hline 
$\Rey$  & 10 & 50 & 80  & 10 & 50 & 80 \\
\hline
$\alpha r_*  = 1 / \Rey^{1/2}$  & 0.31 & 0.14& 0.11 & 0.31 & 0.14& 0.11\\ 
Stresslet with constants (from figure \ref{fig:NSEtermsStresslet})  & 0.9 & 0.4 & 0.3 & 1& 0.3 & 0.2\\
Stresslet with wall effects (from figure \ref{fig:NSEtermsWalleffect}) & $>1$& 0.45 & 0.35 & $>1$& 0.4 & 0.3 \\
NSE with wall effects (from figure \ref{fig:NSEterms})  & $>1$ & 0.6 & 0.4 & $>1$& 0.4 & 0.3 \\[1ex] 
\hline
\end{tabular}
\caption{The cross-over radius $\alpha \, r_* $ at which $I(r)\geq V(r)$ computed for $\Rey =  10, 50, 80$ using the following methods: (i) Ho \& Leal's calculation using the stresslet, (ii) our calculation using the stresslet, (iii) our calculation using the stresslet and first wall correction, and (iv) our calculation using the numerical solution to the full Navier-Stokes equation (NSE). }
\label{tab:oseen}
\end{table}

First we compute the first two terms in the perturbative series $\mathbf{u}^{(0)} + \Rey_p \mathbf{u}^{(1)}$ numerically, showing that retaining these two terms gives the lift force quantitatively accurately over the entire dynamical range of experiments. 

Series expanding (\ref{eq:NSE_diff}) and collecting like terms in $\Rey_p$ we arrive at equations for ($\mathbf{u}^{(0)}$, $p^{(0)}$), the first order velocity and pressure:
\begin{equation}\label{eq:u0}
  \begin{aligned}
  \nabla^2 \mathbf{u}^{(0)} - \nabla p^{(0)} &= \mathbf{0} , \qquad 
  \nabla \cdot \mathbf{u}^{(0)} = 0, \\
  \mathbf{u}^{(0)} &= \mathbf{U_p}^{(0)} + \mathbf{\Omega_p}^{(0)} \times \mathbf{r} - \mathbf{\bar{u}}\mbox{ on }  r = 1, \\
   \mathbf{u}^{(0)} &= \mathbf{0} \mbox{ on channel walls and as } z \to \pm \infty.
   \end{aligned}
\end{equation}
Similarly, the next order velocity and pressure $(\mathbf{u}^{(1)}$, $p^{(1)})$ satisfy the equations:
\begin{eqnarray}
  \nabla^2 \mathbf{u}^{(1)} - \nabla p^{(1)} &=& ( \mathbf{\bar{u}} \cdot \nabla \mathbf{u}^{(0)} + \mathbf{u}^{(0)} \cdot \nabla \mathbf{\bar{u}}  + \mathbf{u}^{(0)} \cdot \nabla \mathbf{u}^{(0)} ) , \qquad
  \nabla \cdot \mathbf{u}^{(1)} = 0, \nonumber \\
  \mathbf{u}^{(1)} &=& \mathbf{U_p}^{(1)} + \mathbf{\Omega_p}^{(1)} \times \mathbf{r} \mbox{ on }  r = 1, \label{eq:u1} \\
   \mathbf{u}^{(1)} &=& \mathbf{0} \mbox{ on channel walls and as } z \to \pm \infty. \nonumber
\end{eqnarray}


For both cases, we only need to solve the Stokes equations with a known body force term. In (\ref{eq:u0}), the body force term is equal to $\mathbf{0}$; in (\ref{eq:u1}) the body force term is equal to the inertia of flow $\mathbf{u}^{(0)}$.  

In fact we can apply Lorentz's reciprocal theorem \citep{Leal80} to calculate the lift force associated with $\mathbf{u}^{(1)}$ without needing to directly solve  (\ref{eq:u1}). We define the test fluid flow $(\hbu,\hp)$ representing Stokes flow around a sphere moving with unit velocity in the $y$-direction: viz satisfying  (\ref{eq:u0}) with the velocity condition on the sphere replaced by $\hbu = \be_y$. If $\bsigma^{(1)}$ and $\hbsigma$ are the viscous stress tensors associated with the flow fields $(\bu^{(1)},p^{(1)})$ and $(\hbu,\hp)$ respectively: $\bsigma^{(1)} = \nabla\bu^{(1)} + (\nabla\bu^{(1)})^T-p^{(1)}\bone$, etc., and $\mathbf{\hat{e}}$ and $\mathbf{e}^{(1)}$ the respective rate-of-strain tensors: $\mathbf{e}^{(1)} = \frac{1}{2}[\nabla\bu^{(1)} + (\nabla\bu^{(1)})^T]$,  etc., then by the divergence theorem, the following relation is valid for any volume $V$ enclosed by a surface $S$.
\begin{equation}
\int_S \left( \bn\cdot\hbsigma\cdot\bu^{(1)}-\bn\cdot\bsigma^{(1)}\cdot\hbu\right) \,{\rm ds} = \int_V \left[ \nabla\cdot\left(\hbsigma\cdot\bu^{(1)}\right)-\nabla\cdot\left(\bsigma^{(1)}\cdot\hbu\right)\right] \,{\rm dv}.
\end{equation}
By setting $V$ equal to the fluid filled domain and substituting boundary conditions from (\ref{eq:NSE_diff}), we deduce:
\begin{equation}\label{eq:RTexpand}
  \begin{split}
    \mathbf{U}_p^{(1)}\cdot \int_S \left(\hbsigma\cdot\bn\right) \,{\rm ds} + \int_S \left( \mathbf{\Omega}_p^{(1)} \times \mathbf{r} \right)\cdot \hbsigma \cdot \bn \,{\rm ds}  - \mathbf{e_y} \cdot \int_S \bsigma^{(1)}\cdot\bn \,{\rm ds} \\ 
    = \int_V \left[ \left(\nabla \cdot \bsigma^{(1)}\right)\cdot \hbu + \bsigma^{(1)}:\mathbf{\hat{e}} -\left(\nabla \cdot \hbsigma\right)\cdot \bu^{(1)} - \hbsigma:\mathbf{e}^{(1)}\right] \,{\rm dv}.
  \end{split}
\end{equation}
On the left hand side of the equation, the first term is zero by symmetry.  Similarly, the integrand of the second term can be rearranged:
\begin{equation}
  \left( \mathbf{\Omega}_p^{(1)} \times \mathbf{r} \right)\cdot \hbsigma \cdot \bn =  \mathbf{\Omega}_p^{(1)}\cdot \left(\mathbf{r} \times \hbsigma\cdot\bn\right),
\end{equation}
which also integrates to zero.  On the right hand side of (\ref{eq:RTexpand}), the third term is zero by definition (since $\hbu$ solves the Stokes equations).  Furthermore, we can rearrange the second and fourth terms:
\begin{equation}
  \bsigma^{(1)}:\mathbf{\hat{e}}  - \hbsigma:\mathbf{e}^{(1)} = 2 \mathbf{e}^{(1)}:\mathbf{\hat{e}} - p^{(1)}\nabla \cdot \hbu - 2 \mathbf{\hat{e}}: \mathbf{e}^{(1)} + \hat{p} \nabla \cdot \bu^{(1)} = 0
\end{equation}
since both flows are incompressible.  So, on the right hand side of (\ref{eq:RTexpand}), only the first term of the volume integral remains.  Using the definitions of $\bsigma^{(1)}$ and $\hbsigma$, we obtain the following formula, which we refer to as the reciprocal theorem.
\begin{equation}\label{eq:recipthm}
  \mathbf{e_y} \cdot \mathbf{f_L} = \int_V  \mathbf{\hat{u}}  \cdot \left(  \mathbf{\bar{u}} \cdot \nabla \mathbf{u}^{(0)} + \mathbf{u}^{(0)} \cdot \nabla \mathbf{\bar{u}}  + \mathbf{u}^{(0)} \cdot \nabla \mathbf{u}^{(0)}  \right)  \mathrm{dv}
\end{equation}

\begin{figure}
  \centerline{
  $(a)$ \includegraphics{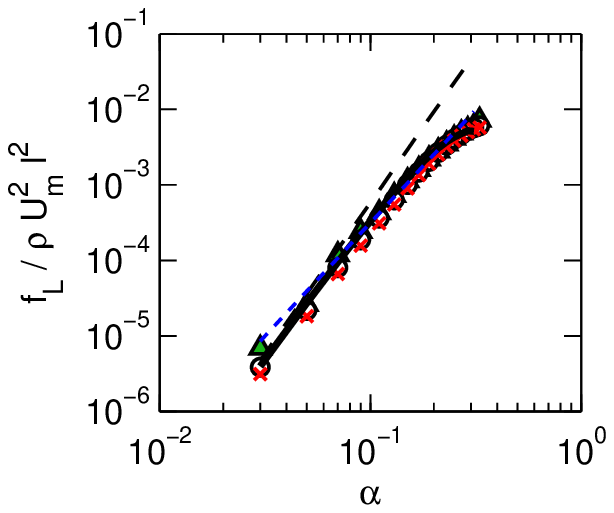}
  $(b)$ \includegraphics{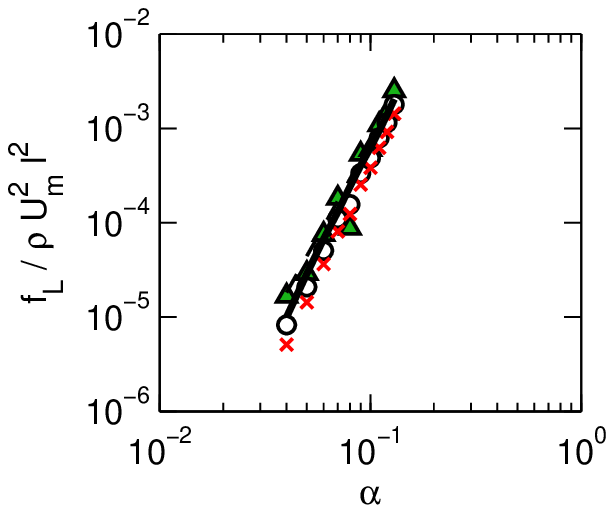}
  }
  \caption{We compute numerically the scaled lift force $f_L/ \rho U_m^2 \ell^2$ using the Navier-Stokes equations in (\ref{eq:NSE_diff}) as a function of particle size $\alpha$ for various channel Reynolds numbers, $\Rey = 10$ (green triangles), $\Rey= 50$ (circles), and $\Rey= 80$ (red x's).  The black dashed line represents a scaling law with exponent 4, i.e. $f_L \sim \rho U_m^2 \alpha^2 a^2$ as in \citet{HoLeal74}, while the dotted blue line represents a scaling law with exponent 3, i.e. $f_L \sim \rho U_m^2 \alpha a^2$, which is the line of best fit computed in \citet{DiCarlo09}.  The solid line represents the regular perturbation expansion computed numerically using the reciprocal theorem in (\ref{eq:recipthm}).  We compare all of these force predictions at two locations in the channel, $(a)$  a particle near the channel center ($y_0=0.15 / \alpha$ and $\alpha=0.11$), and $(b)$ a particle near the channel walls ($y_0=0.35 / \alpha$ and $\alpha=0.06$).}
\label{fig:scalinglaw}
\end{figure}

We have now reduced our calculation of the lift force to that of solving two homogeneous Stokes equations and performing a volume integral. Numerically, we let $V$ be the truncated numerical domain modeled by our FEM. Next we solve numerically for $\mathbf{u}^{(0)}$ from (\ref{eq:u0}) and $\mathbf{\hat{u}}$.  Again, we choose $U_p^{(0)}$ and $\Omega_p^{(0)}$ so that the particle travels force free and torque free.  We compute the lift force using the reciprocal theorem in (\ref{eq:recipthm}) for particles at two different channel positions (figure \ref{fig:scalinglaw}$a$-$b$). We see close quantitative agreement between the lift force computed from the full Navier-Stokes equations and the lift force computed from the reciprocal theorem using the two term expansion in $\Rey_p$. The comparison is accurate even when, as for $y_0 = 0.15 / \alpha$, there is no simple scaling law for the dependence of $f_L$ upon $a$ (figure \ref{fig:scalinglaw}$a$). In the next section, we develop a model that nevertheless allows analytic evaluation of the lift force.

\subsection{Approximation of $\mathbf{u}^{(0)}$ and $\mathbf{\hat{u}}$ by method of images}

In the previous section we showed that a single, regular perturbation in $\Rey_p$ of Stokes equations agrees excellently to the numerically computed lift force. We calculated the terms in this perturbation series numerically but to rationally design inertial microfluidic devices, we need an asymptotic theory for how the lift force and the inertial focusing points depend on the size of the particle and its position within the channel. We derive this theory from asymptotic expansion of $\bu^{(0)}$ and $\hbu$ in powers of $\alpha$, the dimensionless particle size. We follow \cite{HoLeal74} and use the method of reflections to generate expansions in powers of $\alpha$ for the Stokes flow fields appearing in (\ref{eq:recipthm}) \citep{ HappelBrenner83}:
\begin{eqnarray}
  \mathbf{u}^{(0)} &=& \mathbf{u}^{(0)}_1 + \mathbf{u}^{(0)}_2 + \mathbf{u}^{(0)}_3 + \mathbf{u}^{(0)}_4 + \ldots \, ,
\end{eqnarray}
with similar expansions for $p$, $\hbu$, and $\hp$.  Here, $\bu^{(0)}_1$ is the Stokes solution for a particle in unbounded flow,  $\mathbf{u}^{(0)}_2$ is the Stokes solution with boundary condition $  \mathbf{u}^{(0)}_2 = -\mathbf{u}^{(0)}_1$ on the channel walls, and $\bu^{(0)}_3$ is the Stokes solution with boundary condition $\mathbf{u}^{(0)}_3 = -\mathbf{u}^{(0)}_2$ on the particle surface, etc.  Odd terms impose the global boundary conditions on the particle, whereas even terms impose the global boundary conditions on the channel walls.  We will show below that the terms in this series constitute a power series in $\alpha$.

Since the odd terms in the expansion, $\bu^{(0)}_{2i-1}$, are prescribed on the sphere's surface they can be calculated using Lamb's method for solving the flow external to a sphere \citep{Lamb45, HappelBrenner83}. This method expands the velocity field as a sum of multipoles located at the sphere center. Namely,
\begin{equation}
  \bu^{(0)}_{2i-1} = \sum_{n=0}^{\infty} \frac{1}{r^{n+1}} \bfn^i(\frac{x- x_0}{r},\frac{y- y_0}{r},\frac{z}{r}) \, ,
\end{equation}
where each term $\bfn^i / r^{n+1}$ is a combination of the stokeslet $n$-pole and the source $(n-1)$-pole.  
We can similarly expand the odd terms of $\hbu$:
\begin{equation}
  \hbu_{2i-1} = \sum_{n=0}^{\infty} \frac{1}{r^{n+1}} \boldsymbol{\mathrm{g}}_n^i(\frac{x-x_0}{r},\frac{y-y_0}{r},\frac{z}{r}) \, .
\end{equation}
The full analytic forms for the $\bfn^1$ and $\boldsymbol{\mathrm{g}}_n^1$ are listed in Appendix \ref{appVel}. From the analytic form of $\mathbf{u}^{(0)}_1$, we can find $\mathbf{u}^{(0)}_2$ by solving the associated Stokes problem numerically. Given $- \mathbf{u}^{(0)}_2$ on the particle surface, we can appeal to Lamb's solution to find $\mathbf{u}^{(0)}_3$, and so on. The same sequence of reflections can be used to expand the reference velocity $\hbu$.

\subsection{Approximation to the reciprocal theorem integral}

Given the Stokes velocities $\bu^{(0)}$ and $\hbu$ we can compute the inertial lift force $f_L$ up to terms of $O(\Rey_p)$ using the reciprocal theorem (\ref{eq:recipthm}). As in \cite{HoLeal74}, it is advantageous to divide the fluid filled domain $V$ into two subdomains, $V_1$ and $V_2$, where:

\begin{equation}
  V_1 = \{ \mathbf{r} \in  V : r \le \xi \} \quad \hbox{and} \quad V_2 = \{ \mathbf{r} \in V : r \ge \xi\}.
\end{equation}
The intermediate radius $\xi$ is any parameter satisfying $1 \ll \xi \ll \frac{1}{\alpha}$.  Call the corresponding integrals the inner integral and the outer integral, and identify their contributions to the lift force as $f_{L_1}$ and $f_{L_2}$, respectively ($f_L = f_{L_1} + f_{L_2}$).  The division of the integral into inner and outer regions allows one to incorporate varying length scales ($\alpha$ for the inner region and $\ell$ for the outer region) into our model. Note that, distinct from \cite{Hinch89}, inertia remains subdominant even in the outer region $V_2$.  In the next two sections, we will separately consider the contributions from the inner and outer integrals.

\subsection{The Inner Integral}\label{sec:innerint}

For the inner integral we continue to scale lengths by $a$, so that $1\le r \le \xi\ll \alpha^{-1}$.  The inner integral can be expressed as the following expansion in $\alpha$.:
\begin{equation}
  f_{L_1} = \rho U_m^2 a^2 ( h_4 \alpha^2 + h_5 \alpha^3 + \hdots \, )~~.
\end{equation}

In order to calculate the terms $h_4$ and $h_5$, we sort the terms of the Stokes velocities by leading order in $\alpha$. The terms contributing at $O(\alpha^2)$ in the inner region are:
\begin{equation}
 \mathbf{u}^{(0)}_1 \sim  \alpha \left( \frac{1}{r^2}\boldsymbol{\mathrm{f}}_1^1  + \frac{1}{r^4}\boldsymbol{\mathrm{f}}_3^1  \right) , \qquad
 \mathbf{\hat{u}}_1 \sim \frac{1}{r} \boldsymbol{\mathrm{g}}_0^1 + \frac{1}{r^3} \boldsymbol{\mathrm{g}}_2^1\, , \qquad
 \mathbf{\bar{u}} \sim \gamma \alpha r.
\end{equation}
All of these terms are known analytically (see Appendix \ref{appVel}), and it can be shown that their contribution to the inner integral evaluates to zero, i.e. $h_4 = 0$. 

At $O(\alpha^3)$ the velocity terms contributing to calculation of $h_5$ are:

\begin{equation}
\begin{split}
 \mathbf{u}^{(0)}_1 \sim  \alpha \left( \frac{1}{r^2}\boldsymbol{\mathrm{f}}_1^1  + \frac{1}{r^4}\boldsymbol{\mathrm{f}}_3^1  \right)   + \alpha^2 \left( \frac{1}{r^3}\boldsymbol{\mathrm{f}}_2^1  + \frac{1}{r^5} \boldsymbol{\mathrm{f}}_4^1  \right), \, \qquad \mathbf{\bar{u}} \sim \gamma \alpha r + \delta \alpha^2 r^2,  \\
 \mathbf{\hat{u}}_1 \sim  \frac{1}{r} \boldsymbol{\mathrm{g}}_0^1 + \frac{1}{r^3} \boldsymbol{\mathrm{g}}_2^1~, \, \qquad 
 \mathbf{\hat{u}}_2 \sim  \alpha \mathcal{SI}\left[\frac{1}{r}\boldsymbol{\mathrm{g}}_0^1\right]_0, \, \qquad 
 \mathbf{\hat{u}}_3 \sim \alpha \left( \frac{1}{r} \boldsymbol{\mathrm{g}}_0^3 + \frac{1}{r^3} \boldsymbol{\mathrm{g}}_2^3\right) .
\end{split}
\end{equation}
where we define $\mathbf{v}\equiv\mathcal{SI}[\bu]$ as the image of the function $\bu$, and we define $\mathbf{v}_0\equiv\mathcal{SI}[\bu]_0$ as the velocity $\mathbf{v}$ evaluated at the particle center. That is, $\mathbf{v}$ solves the Stokes equations with boundary condition $\mathbf{v}=-\bu$ on the channel walls, and $\mathbf{v}_0=\mathbf{v}( x_0, y_0,0)$.  We determine $\mathcal{SI}\left[\frac{1}{r}\boldsymbol{\mathrm{g}}_0^1 \right]$ numerically, by discretizing Stokes equations as a FEM, with quadratic elements for the velocity field and linear elements for the pressure field, and solving the FEM in Comsol Multiphysics.

The $O(\alpha^3)$ contribution to the inner integral is:
\begin{align}
  h_5 &= \int_{\mathbb{R}^3}  (\mathbf{\hat{u}}_1 + \mathbf{\hat{u}}_2 + \mathbf{\hat{u}}_3 ) \cdot \left(  \mathbf{\bar{u}} \cdot \nabla \mathbf{u}^{(0)}_1 + \mathbf{u}^{(0)}_1 \cdot \nabla \mathbf{\bar{u}}  + \mathbf{u}^{(0)}_1 \cdot \nabla \mathbf{u}^{(0)}_1 \right)  \mathrm{dv}  \nonumber \\
	&= \int_{\mathbb{R}^3}  \mathbf{\hat{u}}_1  \cdot \left(  \mathbf{\bar{u}} \cdot \nabla \mathbf{u}^{(0)}_1 + \mathbf{u}^{(0)}_1 \cdot \nabla \mathbf{\bar{u}}  + \mathbf{u}^{(0)}_1 \cdot \nabla \mathbf{u}^{(0)}_1 \right)  \mathrm{dv}
\end{align}
where we have made use of the fact that the contributions to the integral from $\mathbf{\hat{u}}_2$ and $\mathbf{\hat{u}}_3$ evaluate to zero. Since all of the terms in the integrand are $O(r^3)$ as $r\to \infty$, we can take $\xi \to \infty$; viz, replace integration over $V_1$ by integration over $\mathbb{R}^3$. In doing so, we pick up an error that is $O(1/\xi)$. We neglect this contribution, since $\xi\gg 1$; in fact the error terms can be shown to cancel with corresponding contributions from the outer integral if expansions are continued to higher order powers of $\alpha$. Evaluating the final integral, we obtain:
\begin{equation}
f_{L_1} = \frac{\rho U_m^2 h_5 a^5} { \ell^3} + O(a^6)~,
\end{equation}
where
\begin{equation}\label{eq:h5}
  h_5 = -\frac{26171\pi\gamma_y^2}{277200} - \frac{53\pi\gamma_y\delta_{xx}}{1728}-\frac{283\pi\gamma_y\delta_{yy}}{3150}
\end{equation}
is $O(1)$, and depends only on the location of the particle.  Recall that the constants $\gamma_y, \delta_{xx}$, and $\delta_{yy}$ were defined in the expansion of $\bar{u}$ in (\ref{eq:ubarseries}), and depend on the particle position.

\subsection{The Outer Integral}\label{sec:outerint}

For the outer integral we will consider alternate dimensionless variables, by using the rescaled distance $\mathbf{R} = \alpha \mathbf{r}$. This corresponds to using $\ell$ to non-dimensionalize lengths, rather than $a$.  We call these variables the outer variables, and we will denote them with uppercase roman letters.  A detailed comparison of the dimensionless variables is given in Appendix \ref{appNotation}.
 
In the outer region $V_2$, we must express our functions in terms of $\mathbf{R}$ and rearrange our functions by order of magnitude in $\alpha$.  These expansions are listed in full in Appendix \ref{appIntegrand}. 
In the outer region, the reciprocal theorem integral takes the following dimensional form:
\begin{equation}
  f_{L_2}= \rho U_m^2 \ell^2 \int_{V_C}  \mathbf{\hat{U}}  \cdot \left(  \mathbf{\bar{U}} \cdot \nabla \mathbf{U}^{(0)} + \mathbf{U}^{(0)} \cdot \nabla \mathbf{\bar{U}}  + \mathbf{U}^{(0)} \cdot \nabla \mathbf{U}^{(0)}  \right)  \mathrm{dv},
\end{equation}
where we have expanded our domain of integration from $V_2= \{ \mathbf{R} \in V : R \ge \xi\}$ to the entire empty channel $V_C$.  This expansion of the domain is justified since the contribution from the region that we add to the integral  $\{ \mathbf{R} : 0 \le R \le \alpha \xi\}$ is $O(\alpha^4 \xi)$, and $\xi \ll 1/\alpha$. In fact this residue (which would show up in the $O(\alpha^3)$ inner integral) is exactly zero.

As we did for the inner integral, we can write the outer integral as an expansion in $\alpha$.
\begin{equation}
  f_{L_2} = \rho U_m^2 \ell^2 ( k_4 \alpha^4 + k_5 \alpha^5 + \hdots \,) ~.
\end{equation}

The velocity terms that contribute to $k_4$ are the following:

\begin{equation}
\begin{split}
  \mathbf{U}^{(0)}_1 \sim \alpha^3 \frac{1}{R^2} \boldsymbol{\mathrm{f}}_1^1~,  \, \qquad
  \mathbf{U}^{(0)}_2 \sim  \alpha^3 \mathcal{SI}\left[\frac{1}{R^2}\boldsymbol{\mathrm{f}}_1^1\right], \\
  \mathbf{\hat{U}}_1 \sim \alpha \frac{1}{R} \boldsymbol{\mathrm{g}}_0^1~, \, \qquad
	\mathbf{\hat{U}}_2 \sim \alpha \mathcal{SI}\left[\frac{1}{R}\boldsymbol{\mathrm{g}}_0^1\right],  \, \qquad 
	\mathbf{\bar{U}} \sim \gamma R + \delta R^2 + \ldots~~.
\end{split}
\end{equation}
Again, we define $\mathbf{V}=\mathcal{SI}[\mathbf{U}]$ as the image of the function $\mathbf{U}$, and we compute $\mathcal{SI}\left[\frac{1}{R^2}\boldsymbol{\mathrm{f}}_1^1\right] $ and $\mathcal{SI}\left[\frac{1}{R}\boldsymbol{\mathrm{g}}_0^1\right]$ numerically.  Furthermore, we can approximate the term $\boldsymbol{\mathrm{f}}_1^1$ by the stresslet terms, since the rotlet terms have coefficients that are order $O(\alpha^2)$ higher than the coefficients of the stresslet terms. The $O(a^4)$ contribution to the reciprocal theorem integral takes the following form:
\begin{equation}\label{eq:outer}
  k_4 = \int_{V_C} ( \mathbf{\hat{U}}_1 + \mathbf{\hat{U}}_2 )  \cdot \left[  \mathbf{\bar{U}} \cdot \nabla (\mathbf{U}^{(0)}_1 + \mathbf{U}^{(0)}_2) + (\mathbf{U}^{(0)}_1 + \mathbf{U}^{(0)}_2) \cdot \nabla \mathbf{\bar{U}}   \right]  \mathrm{dv} .
\end{equation}

We run into a problem numerically evaluating the integral in (\ref{eq:outer}) when considering only the first terms in the series expansions, $\mathbf{U}^{(0)}_1$ and $\mathbf{\hat{U}}_1$. The problem arises because $\mathbf{U}^{(0)}_1$ and $\mathbf{\hat{U}}_1$ have singularities of the form:
\begin{equation}
  \mathbf{U}^{(0)}_1  \approx - \frac{5\gamma_y}{2} \frac{(Y-Y_0)Z\mathbf{R}}{R^5} ~, \qquad
  \mathbf{\hat{U}}_1 \approx \frac{3}{4} \left(\mathbf{e}_Y+\frac{(Y-Y_0)\mathbf{R}}{R^2}\right)\frac{1}{R} ~,
\end{equation}
which are respectively the stresslet and stokeslet components of the two velocity fields.  When the singularities are integrated against the shear term of $\mathbf{\bar{U}}$, that is  $\mathbf{\bar{U}}_\gamma \approx \gamma_y Y \,\mathbf{e_Z}$, the result is an integral that is undefined near $R=0$.

\begin{equation}\label{eq:singshear}
      \int_{R<\epsilon} \mathbf{\hat{U}}_1  \cdot \left[  \mathbf{\bar{U}}_\gamma  \cdot \nabla \mathbf{U}^{(0)}_1  + \mathbf{U}^{(0)}_1 \cdot \nabla \mathbf{\bar{U}}_\gamma   \right]  \mathrm{dv}
\end{equation}

\noindent However, converting to spherical coordinates, we find that the angular dependence forces the integral in (\ref{eq:singshear}) to be zero:

\begin{equation}
    \negthickspace \iiint\limits_{0~0~0}^{\quad \pi~2\pi~\epsilon}  \left(\frac{15 \gamma_y^2 (1+2\cos{2\theta}) \sin^4{\theta} \sin^3{\phi}}{4R}\right) ~\mathrm{d}R \mathrm{d}\phi \mathrm{d}\theta =0.
\end{equation}
This angular behavior is difficult to capture numerically, especially if the mesh is not symmetric.  Instead, we propose a regularization of the outer integral, where we integrate the problematic terms analytically in a small region near $R=0$.  Now considering the full expansion of $\bar{u}$, we derive the following analytic form for the integral in the region near the origin:

\begin{equation}
   \int_{R<\epsilon}\mathbf{\hat{U}}_1  \cdot \left[  \mathbf{\bar{U}} \cdot \nabla \mathbf{U}^{(0)}_1  + \mathbf{U}^{(0)}_1 \cdot \nabla \mathbf{\bar{U}}   \right]  \mathrm{dv} = - ~\pi \gamma_y (\delta_{xx}+3\delta_{yy}) \epsilon
  \end{equation}

Recall that the constants $\gamma_y$, $\delta_{xx}$, and $\delta_{yy}$ were defined in the expansion of $\mathbf{\bar{u}}$ in (\ref{eq:ubarseries}).  Using this analytic expression, we split up the rest of the reciprocal theorem integral (\ref{eq:outer}) into the following parts.

\begin{eqnarray}\label{eq:regularize}
    k_4&=&  \int_{V_C}  \mathbf{\hat{U}}_2  \cdot \left[  \mathbf{\bar{U}} \cdot \nabla (\mathbf{U}^{(0)}_1 + \mathbf{U}^{(0)}_2) + (\mathbf{U}^{(0)}_1 + \mathbf{U}^{(0)}_2) \cdot \nabla \mathbf{\bar{U}}   \right]  \mathrm{dv} \nonumber \\ 
    &+&  \int_{V_C}  \mathbf{\hat{U}}_1  \cdot \left[  \mathbf{\bar{U}} \cdot \nabla  \mathbf{U}^{(0)}_2 + \mathbf{U}^{(0)}_2 \cdot \nabla \mathbf{\bar{U}}   \right]  \mathrm{dv} \\
    &+&  \int_{\{\mathbf{r} \in V_C :R\ge\epsilon\}} \mathbf{\hat{U}}_1  \cdot \left[  \mathbf{\bar{U}} \cdot \nabla \mathbf{U}^{(0)}_1  + \mathbf{U}^{(0)}_1 \cdot \nabla \mathbf{\bar{U}}   \right]  \mathrm{dv} \nonumber \\
    &-&  ~\pi \gamma_y (\delta_{xx}+3\delta_{yy}) \epsilon ~. \phantom{\int_V \mathbf{\hat{U}} dv} \nonumber
\end{eqnarray}

The first three lines in (\ref{eq:regularize}) are evaluated numerically using the FEM. Evaluating the integral in (\ref{eq:regularize}), we arrive at the scaling law:
\begin{equation}
  f_{L_2} = \frac{\rho U_m^2 k_4 a^4 }{\ell^2} + O(a^5)~,
\end{equation}
where $k_4 = O(1)$ is a constant that depends on the location of the particle in the channel, and is computed numerically.

Similarly, the $O(\alpha^5)$ correction to the outer integral comes from terms:

\begin{equation}
\begin{split}
  \mathbf{U}^{(0)}_1 \sim \alpha^3 \frac{1}{R^2} \boldsymbol{\mathrm{f}}_1^1 ~, \qquad 
   \mathbf{U}^{(0)}_2 \sim  \alpha^3 \mathcal{SI}\left[\frac{1}{R^2}\boldsymbol{\mathrm{f}}_1^1\right],   \\
   \mathbf{\hat{U}}_3 \sim \alpha^2 \frac{1}{R} \boldsymbol{\mathrm{g}}_0^3~,  \qquad 
    \mathbf{\hat{U}}_4 \sim \alpha^2 \mathcal{SI}\left[\frac{1}{R}\boldsymbol{\mathrm{g}}_0^3\right]  , \quad
	\mathbf{\bar{U}} \sim \gamma R + \delta R^2 + \ldots ~~.
\end{split}
\end{equation}

\noindent Again, we must regularize the outer integral, since $\mathbf{\hat{U}}_3$ also has a stokeslet singularity.  We use the same regularization as before, replacing $\mathbf{\hat{U}}_1$ and $\mathbf{\hat{U}}_2$ with $\mathbf{\hat{U}}_3$ and $\mathbf{\hat{U}}_4$, respectively.

And, combining terms at $O(a^4)$ and $O(a^5)$, we obtain:
\begin{equation}\label{eq:fl2_outer}
  f_{L_2} = \frac{\rho U_m^2 k_4 a^4}{ \ell^2} + \frac{\rho U_m^2 k_5 a^5} {\ell^3} +O(a^6)~,
\end{equation}
where $k_5 = O(1)$ is a constant that depends on the location of the particle in the channel.  We have now calculated the $V_2$ contribution to the reciprocal theorem integral up to order $O(a^5)$.

\subsection{Results}\label{sec:results}

\begin{figure}
  \centerline{
  $(a)$ \includegraphics{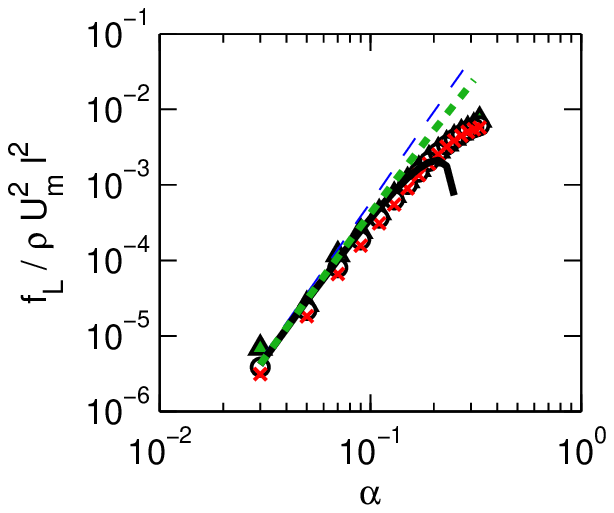}
  $(b)$ \includegraphics{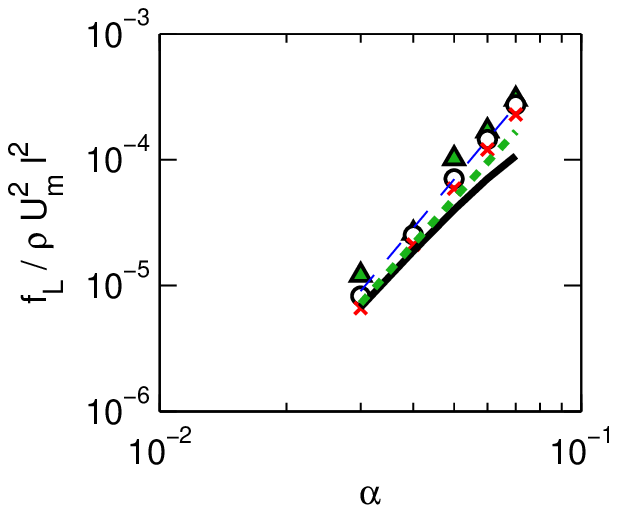}
  }
  \caption{We compute lift force $f_L$ numerically using the Navier-Stokes equations in (\ref{eq:NSE_diff}) and plot as a function of particle radius $a$ for channel Reynolds numbers $\Rey = 10$ (triangles), $\Rey= 50$ (circles), and $\Rey= 80$ (x's).  The blue dashed line represents a scaling law of particle radius to the fourth power $f_L = \rho U_m^2 c_4 \alpha^2 a^2$, the solid black line represents the sum of the fourth and fifth power terms in (\ref{eq:2dFL}), and the green dotted line represents the completion of series in (\ref{eq:complseries}), with $(a)$ particle displacement $ y_0=0.15 / \alpha$ , and $(b)$ particle displacement $y_0=0.4 / \alpha$.}
\label{fig:order5}
\end{figure}

In the last section, we described our method of computing the correction to the scaling law made by \cite{HoLeal74}.  Combining the inner and outer integrals, the result is a new approximation of the form:
\begin{equation}\label{eq:2dFL}
  f_L = \frac{\rho U_m^2 c_4 a^4}{\ell^2} + \frac{\rho U_m^2 c_5 a^5}{\ell^3} + O(a^6) \, ,
\end{equation}
where $c_4 = k_4$ from (\ref{eq:regularize}), and $c_5 = h_5+k_5$ from (\ref{eq:h5}) and (\ref{eq:fl2_outer}). The prefactors $c_4$ and $c_5$ are $O(1)$ in $\alpha$, and depend only on the location of the particle in the channel. The extended series agrees well with numerical data for particle sizes up to $\alpha=0.2-0.3$ (Fig. 6). This calculation could in principle be extended by computing the contributions from higher order terms. Completing the series \citep{ Hinch91}, i.e. approximating:
\begin{equation}\label{eq:complseries}
f_L \approx \frac{\rho U_m^2 c_4 a^4}{\ell^2\left(1-\frac{c_5 a}{c_4\ell}\right)}~,
\end{equation}
produces a modest increase in the accuracy of the asymptotic approximation (Fig. 6).

By including two terms in our asymptotic expansion, we can describe how the particle equilibrium position depends on its size -- a key prediction for rationally designing devices that use inertial lift forces to fractionate particles, or to transfer them between fluid streams \citep{DiCarlo09,DiCarlo10,DiCarlo11,DiCarlo13,DiCarlo14} (figure \ref{fig:liftforce_channel}). We compare our asymptotic calculation predictions directly with experiments of \citet{DiCarlo09}, finding good agreement in focusing positions up to $a=0.3$ (Fig. \ref{fig:liftforce_channel}$b$.).

\section{3D asymptotic expansion}\label{sec:3d}

Previous asymptotic studies have considered inertial migration in 2D flows \citep{HoLeal74,Hinch89,Hogg94, Asmolov99}.  At sufficiently small values of $a$ there is qualitative agreement between the 2D theories and our theory, but only when the particle is located on a symmetry plane e.g. $x_0=0$ or $y_0=0$.  However, real inertial microfluidic devices focus in $x$ and $y$- directions, taking initially uniformly dispersed particles to four focusing positions. Our asymptotic approach allows us to compute the focusing forces for particles placed at arbitrary positions in the channel.

The calculation is very similar to the one outlined in \S \ref{sec:asymptotics}; we only need to add similar terms driven by the shear in the $x$-direction, and allow for a reciprocal velocity $\hat{\boldsymbol{u}}$ associated with moving the particle in this direction.  The full Lamb's solution for $\mathbf{u}^{(0)}_1$ has additional terms from the shear in the x-direction (i.e. the terms with coefficients $\gamma_x$), shown in Appendix \ref{appVel}.  The only additional components of $\bu^{(0)}_1$ that contribute to the 3D calculation are the stresslet and source quadrupole.
%

\begin{figure}
  \centerline{
  $(a)$ \includegraphics{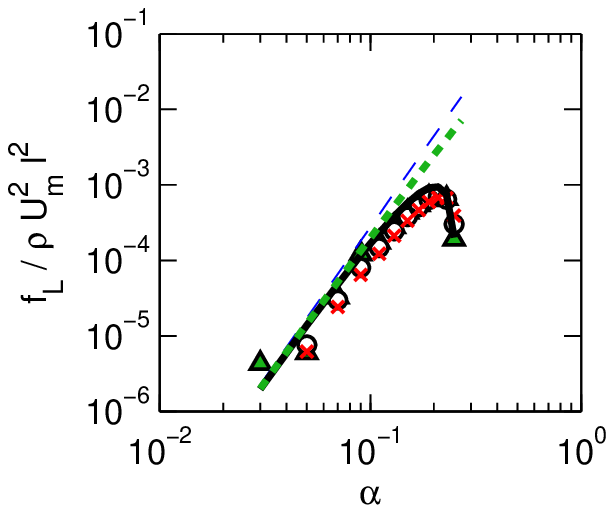}
  $(b)$ \includegraphics{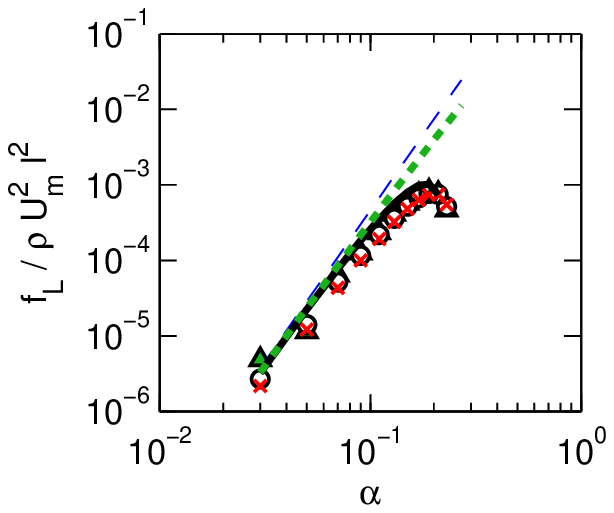}
  }
  \caption{We compute lift force $f_L$ numerically from (\ref{eq:NSE_diff}) as function of particle radius for channel Reynolds numbers $\Rey = 10$ (triangles), $\Rey= 50$ (circles), and $\Rey= 80$ (x's).  The blue dashed line represents a scaling law of particle radius to the fourth power $f_L = \rho U_m^2 c_4 \alpha^2 a^2$, the solid black line represents the fifth power correction term in (\ref{eq:3dFL}), and the green dotted line represents the completion of series in (\ref{eq:complseries}), with particle displacement $x_0=0.2 / \alpha$ and $y_0=0.15 / \alpha$ for $(a)$ lift force in the x-direction and $(b)$ in the y-direction.}
\label{fig:order53D}
\end{figure}

\begin{figure}
  \centerline{
  $(a)$\includegraphics{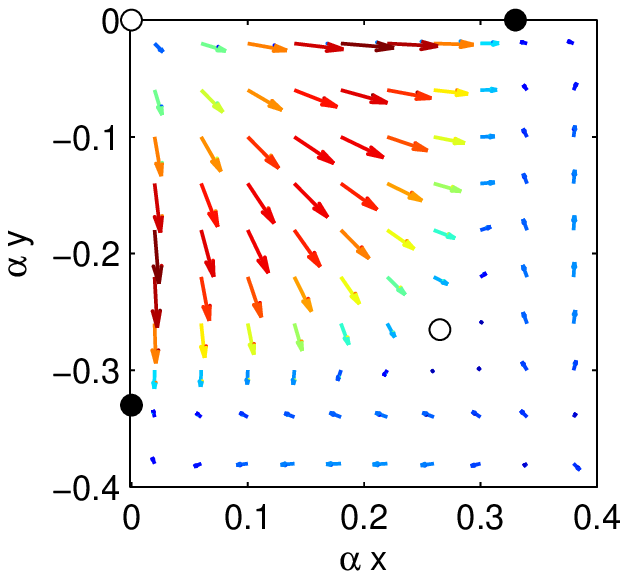}
  $(b)$\includegraphics{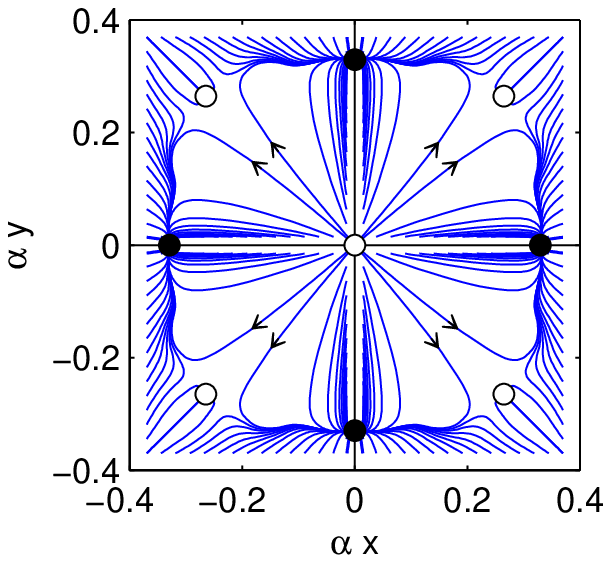}
  }
  \caption{$(a)$ Lift force calculated using (\ref{eq:3dFL}) for locations in the lower right quadrant of the channel for a particle of radius $\alpha=0.11$, and $\Rey=80$. The solid black circles mark stable equilibrium points, while the open white circles mark unstable equilibrium points.  $(b)$ Trajectories of particles calculated using (\ref{eq:trajectory}) for particle size $\alpha=0.11$ and $\Rey=80$. The solid black circles mark stable equilibrium points, while the open white circles mark unstable equilibrium points. }
\label{fig:liftforce_channel}
\end{figure}

The inner integral in 3D evaluates to:
\begin{equation}
f_{L_1}^{(3D)} = \frac{\rho U_m^2  h_5^{(3D)} a^5} { \ell^3} + O(a^6)~,
\end{equation}
where
\begin{eqnarray}
  h_5^{(3D)} &=& \frac{4381\pi\gamma_x \gamma_y}{554400} -\frac{26171\pi\gamma_y^2}{277200} + \frac{527 \pi \psi_y \gamma_x \gamma_y}{116424} - \frac{53\pi\gamma_y\delta_{xx}}{1728} \\
  &+& \frac{19\pi \gamma_x\delta_{yy}}{3150}  -\frac{283\pi\gamma_y\delta_{yy}}{3150} .\nonumber
\end{eqnarray}
We define $\psi_y$ to be the value of y-component of image of the stokeslet evaluated at the location of the particle:

\begin{equation}\label{eq:psiy}
  \psi_y = \left[\mathcal{SI}\left[  \frac{1}{r} \boldsymbol{\mathrm{g}}_0^1  \right] \cdot \mathbf{e_y} \right]\Bigg|_{(x,y,z)=( x_0, y_0,0)},
\end{equation}
where $\boldsymbol{\mathrm{g}}_0^1$ is the stokeslet and the leading term of $\hbu_1$ defined in (\ref{eq:g01}) in Appendix \ref{appVel}.  The outer integral remains the same, however, $\mathbf{u}^{(0)}_1$ and $\mathbf{u}^{(0)}_3$ each now include a stresslet contribution associated with shear in the x-direction.  Computing this integral gives a scaling law of the form:
\begin{equation}\label{eq:3dFL}
  f_L ^{(3D)}=  \frac{\rho U_m^2 c_4^{(3D)} a^4}{ \ell^2} + \frac{\rho U_m^2 c_5 ^{(3D)} a^5}{\ell^3} + O(a^6)~.
\end{equation}
It remains true that for particles located arbitrarily in the square channel, the lift force scales like $a^4$ in the asymptotic limit $\alpha \to 0$.  Additionally, our $O(a^5)$ correction to the scaling law remains accurate for moderately large $\alpha$, shown in figure \ref{fig:order53D}$a$ and \ref{fig:order53D}$b$ for the forces in the $x$ and $y$-direction, respectively.  We provide the calculated values of the three dimensional lift force in a square channel in a Matlab code in the online supplementary materials.  In particular we find that lift forces vanish only at 8 symmetrically placed points around the channel, with 4 points being stable and 4 unstable, in good agreement with experimental observations (Fig. \ref{fig:liftforce_channel}$a$).  

We can compute particle streamlines using the lift force prediction, and confirm that there are four stable focusing positions in the channel (figure \ref{fig:liftforce_channel}$b$).  Particles are advected using a Forward Euler time stepping scheme.  We find the particle velocity by equating the $O(a^5)$ lift force (\ref{eq:3dFL}) with the $O(a)$ drag force \citep{HappelBrenner83}.  That is, $v_L$, the y-component of velocity $v$, satisfies the equation:
\begin{equation}\label{eq:trajectory}
  6\pi\mu a (v_L + \psi_y)= \left[\frac{\rho U_m^2 c_4^{(3D)} a^4}{ \ell^2} + \frac{\rho U_m^2 c_5 ^{(3D)} a^5}{\ell^3}\right] \Bigg|_{(x_0,y_0,0)} \, ,
\end{equation}
where $\psi_y$ is the image velocity of the stokeslet defined in (\ref{eq:psiy}).  The velocity $u_L$, the x-component of velocity $u$, is computed in the same way by substituting $\psi_x$ from the x-stokeslet for $\psi_y$.

\begin{figure}
  \centerline{
   $(a)$ \includegraphics{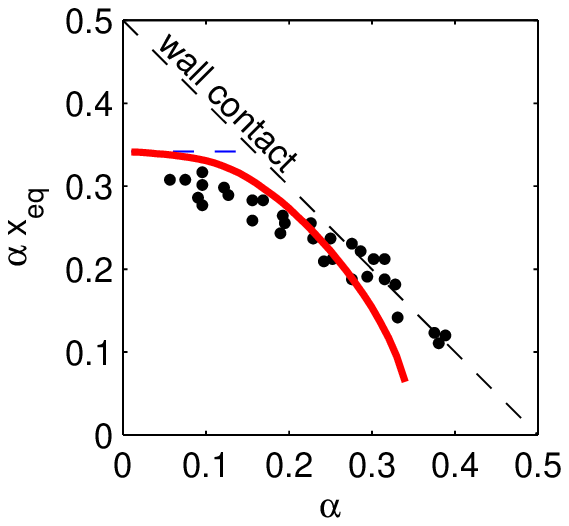}
   $(b)$ \includegraphics{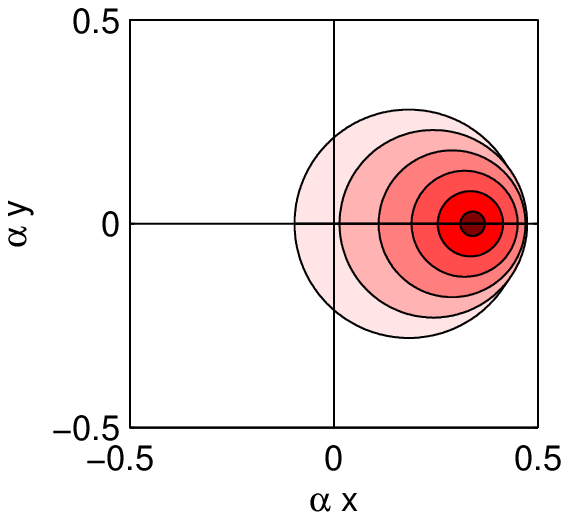}
  }
  \caption{$(a)$ Our theory predicts inertial focusing position as a function of particle size. The markers are data collected by \cite{DiCarlo09}, the dashed blue line is the theory predicted by the first term of $O(a^4)$ in (\ref{eq:2dFL}), and the solid red line is the theory predicted by (\ref{eq:2dFL}). $(b)$ In this schematic diagram we plot the outlines of particles at their predicted focusing position along the positive x-axis. The particle sizes range between $\alpha = 0.03$ and $\alpha = 0.29$.  }
\label{fig:eqpoints}
\end{figure}

In addition, the distance of the focusing positions from the channel center-line can be predicted by solving the implicit equation $f_L^{(3D)} = 0$.   Recall that the lift force coefficients depend on the location of the particle, i.e. $c_4^{(3D)}=c_4^{(3D)}(x_0,y_0)$ and $c_5^{(3D)}=c_5^{(3D)}(x_0,y_0)$.  Since the lift force formula has both $O(a^4)$ and $O(a^5)$ terms, the focusing position will have a functional dependence on the particle size $a$.  This prediction of the focusing position compares well with experimental data by \citet{DiCarlo09}, especially for particle sizes up to $\alpha \le 0.3$ (Fig. \ref{fig:eqpoints}$b$).

\section{Discussion}\label{sec:discussion}

Our findings resolve confusion about the size dependence of inertial lift forces experienced by particles traveling through microchannels. Many asymptotic and numerical studies have been employed to determine how the lateral force scales with particle radius, and have found power laws with exponents two, three, four, and five.  By numerically dissecting the equations of fluid flow around the particle, we find that viscous stresses dominate over inertial stresses even at moderate channel Reynolds numbers. We rationalize this finding by showing that this ordering of fluxes is inherited from the stresslet flow field approximation to the far field of a particle, provided that the contribution from channel walls is included. We make use of this fact to develop a perturbation series expansion for the lift force, extending the theory of Ho \& Leal both to three dimensions and to include $O(a^5)$ sized terms. We find that the scaling is a power law with exponent four for asymptotically small particle radius, but that additional terms must be included to predict lift forces for the range of particle sizes and flow speeds accessed in real inertial microfluidic devices. By including these additional terms, we are also able to predict asymptotically how focusing position depends on particle size.

Somewhat surprisingly, the regular perturbation expansion accurately predicts the particle lift force even at channel Reynolds numbers and particle sizes where the parameters in our expansion are not small (e.g. up to $\Rey_p \approx 10$). This is consistent with our determination that inertial stresses fluxes scale simply with $U^2$ even outside of the regime of velocities and channel sizes at which viscous stresses are numerically larger than momentum fluxes. Thus although assuming a viscous stress-pressure dominant balance is not justified based on simple comparison of the order of magnitude of terms, the perturbation expansion continues to give good results.

We hope that the results in this paper will provide a first step toward predictive theory for the design of inertial microfluidic devices. The biggest unmet challenge here is to determine whether unsteady effects scale like momentum fluxes for determining dominant balances. If the unsteady scaling can be established, then it will be possible to model time varying problems, including the migration of particles in non-rectilinear geometries, such as the microcentrifuge, or the interactions of particles, such as the recently discovered phenomena of self-organization by inertially focused particles into stably ordered chains \citep{lee2010dynamic,humphry2010axial}. We have shown that the viscous-pressure stress dominant balance leads to a particularly simple far-field form to the flow disturbance, potentially allowing simplified modeling of particle interactions.  Additionally we provide a Matlab code with the calculated values of the lift force in the online supplementary materials.

This work was partly supported by the National Science Foundation through grants DGE-1144087 (to K. Hood) and DMS-1312543 (to M. Roper) and by a research fellowship from the Alfred P. Sloan Foundation to M. Roper. We thank Dino DiCarlo, Howard Stone and Z. Jane Wang for useful discussions.

\appendix

\section{Notation}\label{appNotation}

\begin{table}
\centering 
\begin{tabular}{llll}
Variables & Dimensional  &  Inner  &  Outer   \\ \hline
Distance & $\mathbf{r}' = (x', y', z')$ & $\mathbf{r} = (x, y, z)$ & $\mathbf{R} = (X,Y,Z)$ \\
Velocity & $\mathbf{u}'$ & $\mathbf{u}$ & $\mathbf{U}$ \\
Pressure & $p'$ & $p$ & $P$ \\ \hline
Particle location & $(x_0', y_0')$ & $(x_0, y_0)$ & $(X_0, Y_0)$ \\ 
Particle velocity & $\mathbf{U}_p'$ & $ \mathbf{U}_p$ & $\mathbf{U}_p$ \\ 
Particle angular velocity & $\mathbf{\Omega}_p'$ & $\mathbf{\Omega}_p$ & $\mathbf{\Omega}_p$\\ \hline
Poiseuille flow & $\mathbf{\bar{u}}'$ & $\mathbf{\bar{u}}$ & $\mathbf{\bar{U}}$ \\
Asymptotic expansion of velocity & - & $\mathbf{u}^{(0)}$ & $\mathbf{U}^{(0)}$ \\
Reference velocity & - & $\mathbf{\hat{u}}$ & $\mathbf{\hat{U}}$ \\ \hline
Conversion from Dimensional variables & $\mathbf{r}'$ & $\mathbf{r} = \mathbf{r}' /a$ & $\mathbf{R} = \mathbf{r}' / \ell$ \\
Conversion from Inner variables & $\mathbf{r}' = a \mathbf{r}$ & $\mathbf{r}$ & $\mathbf{R} = \alpha \mathbf{r}$\\
Conversion from Outer variables & $\mathbf{r}' = \ell \mathbf{R}$ & $\mathbf{r} = \mathbf{R} / \alpha$ & $\mathbf{R}$ \\
\hline
\end{tabular}
\caption{Comparison of dimensional and dimensionless scalings of the variables.}
\label{tab:vars}
\end{table}

Throughout the main paper, we need to change the scaling of variables in order to capture the dynamics either near the particle or near the channel walls. In this Appendix we create a reference for the notation for three scalings: dimensional scalings, dimensionless inner variables, and dimensionless outer variables.  To be consistent with previous literature, we denote the dimensionless inner variables with lower case roman letters, and the dimensionless outer variables with upper case roman letters.  We use primes to distinguish dimensional variables.  A reference is presented in table \ref{tab:vars}.

We must draw attention to the notation for the particle velocity and particle angular velocity.  Since both the inner and outer variables are scaled by the same velocity, $\alpha U_m$, the scaled particle velocity is the same in both cases.  We choose to represent the dimensionless particle velocity by $\mathbf{U}_p$ to be consistent with notation in previous studies \citep{Hinch89, HoLeal74}.

However, the scaling for the particle angular velocity differs between the inner and outer coordinates.  We continue to use $\mathbf{\Omega}_p$ in both the inner and outer variables, despite this abuse of notation. We keep $\mathbf{\Omega}_p$ in order to be consistent with previous studies \citep{Hinch89, HoLeal74}.  We feel justified in our decision since $\mathbf{\Omega}_p$ does not arise in the computation of our asymptotic model, so the reader wishing to apply our results need not worry over the discrepancy.  Nevertheless, we remind the reader that the particle angular velocity for the inner variables satisfies $\mathbf{\omega}_p = \alpha \mathbf{\Omega}_p$.

\section{Accuracy of the numerical model}\label{appModel}

\subsection{Accuracy of the domain size}

We subjected the FEM discretization of (\ref{eq:NSE_diff}) to convergence tests based on varying the size of the numerical domain and on changing the mesh size. Maximum element size was decreased, and the length of the domain was increased until the computed drag and lift forces had converged to within 0.5\%. 

To test the length of the domain, we varied the variable $L_z$, defined so that the channel domain became $ \frac{\ell}{a} \times \frac{\ell}{a} \times L_z \frac{\ell}{a}$.  We solved the Navier-Stokes equations where the particle surface and channel walls have no slip boundary conditions.  The particle is assumed to have no velocity or angular velocity. Comsol's standard meshing algorithms are used, and the particle size is chosen to be $\alpha = 0.11$.

\begin{figure}
  \centerline{
     $(a)$ \includegraphics{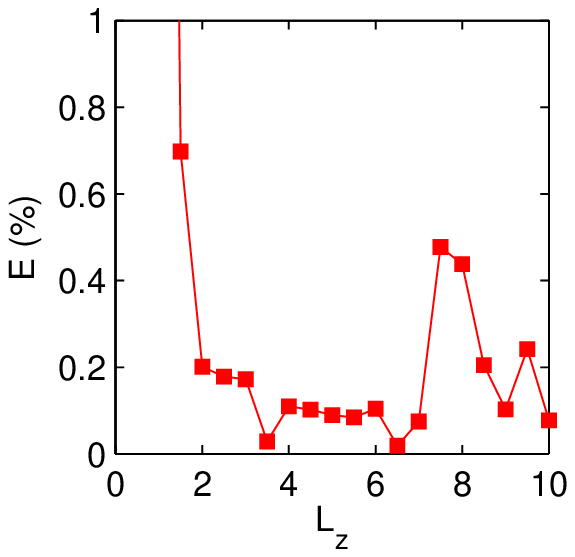} 
     $(b)$ \includegraphics{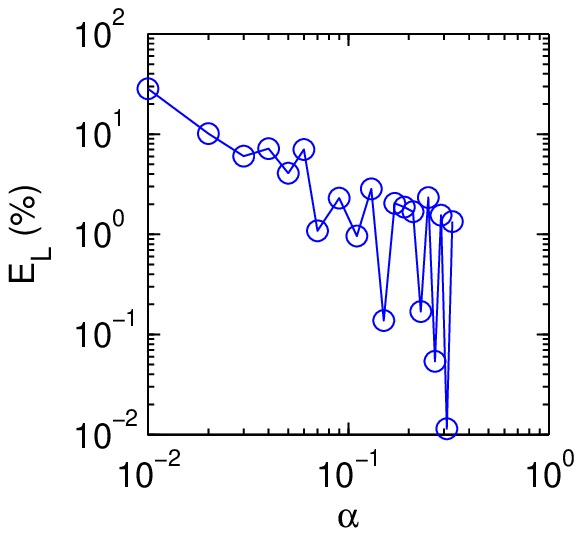} 
   } \caption{$(a)$ The relative error $E$ of the drag force is less than 1\% for $L_z>1$, and in particular for our choice of $L_z=5$ the relative error is less than 0.5\%.  $(b)$ The relative lift force error $E_L$ from solving the Navier-Stokes equations increases exponentially as $\alpha$ decreases. }
\label{fig:Lz}
\end{figure}

We see that the drag force $F$ quickly converges to its final value, at about $L_z = 2-3$  (figure \ref{fig:Lz}$a$). Averaging the data for $L_z>=3$ to obtain a force estimate $\bar{F}$, we also present a relative error as $E = 100 (F - \bar{F})/\bar{F}$. A small fluctuating error persists as $L_z$ is increased up to $L_z=10$. This error probably reflects mesh noise, rather than geometry. Similar data is seen when simulations are performed on different grades of mesh.

\subsection{Accuracy of the particle size}

We also discuss the range of $\alpha$ that were computed in this paper.  The upper range of $\alpha$ is limited by the position of the particle relative to the walls and to the equilibrium positions.  Since the lift force $f_L$ is not only a function of $\alpha$, but also of $x_0$ and $y_0$, the location of the particle in the channel.  For any given $\alpha$, there are four coordinates $(x_0, y_0)$ where the lift force is zero, we call these points equilibrium positions.  In the first part of our paper, we constrain our locations to those on the positive y-axis, that is coordinates of the form $(0,y_0)$, for $y_0>0$.  The equilibrium position $(0,y^*_0)$, which is a function of $\alpha$, divides this domain into two sections: (i) domain between the equilibrium position and the center of the channel, and (ii) the domain between the equilibrium position and the wall.  We must be careful in these regions, when we are examining the scaling law for fixed $y_0$ and varied $\alpha$, to choose $\alpha$ so that $f_L$ remains positive and does not pass through zero.  The same principal extends to the choice of $\alpha$ for all locations throughout the channel.

The lower range of $\alpha$ however, is limited by the accuracy of the root finder in the Navier-Stokes solver.  As discussed in \S\ref{sec:eqns}, the values of $U_p$ and $\mathbf{\Omega}_p$ are chosen so that the particle is drag free and force free.  Therefore, the drag force $F_D$ can be used as a measure of the precision of the numerical solver.  We define the relative lift force error as $E_L = 100(F_D / f_L)$.  The relative lift force error as a function of $\alpha$ for an particle located at $(x_0=0, y_0=0.15/\alpha)$ and with $\Rey = 1$ increases exponentially as $\alpha$ decreases to zero (figure \ref{fig:Lz}$b$).  We limit $\alpha >= 0.03$, or relative error $E_L < 10\%$.

We do not strive to test smaller particles sizes $\alpha$ because the theoretical results of \cite{Hinch89} and \cite{Asmolov99} are generally accepted to be true for asymptotically small particles. Our goal in this paper is to produce theory for particles at the sizes used in experimental systems.  Smaller values of $\alpha$ are not used in experiments, because lift forces become too weak to compete with other forces, such as Brownian motion.

\subsection{Accuracy of the Naver-Stokes solver}

\begin{figure}
  \centerline{
        \includegraphics{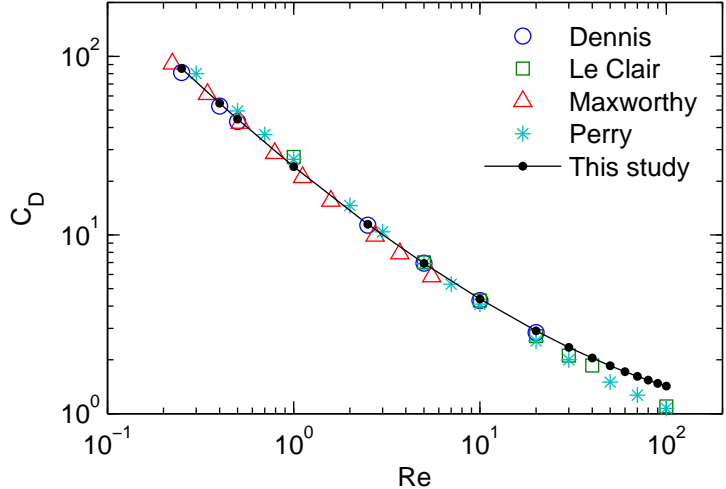} 
   } \caption{ Our calculation of the drag coefficient, $C_D$, for a sphere in a uniform flow (black dot) compares well to numerical data (Dennis - blue circles, Le Clair - green squares) and experimental data (Maxworthy - red triangles, Perry - cyan *) across a large range of Reynolds numbers, $\Rey$.}
\label{fig:dragvRe}
\end{figure}

To further account for artifacts associated e.g. with regularization of the convective (inertial) terms in (\ref{eq:NSE_diff}), we also solved a model problem of computing the drag force on a sphere moving through a quiescent fluid, for which a considerable body of well-validated experimental and numerical data exists \citep{Goldenfeld2007}.  

Let $a$ be the particle radius, $U\mathbf{e_z}$ be the flow velocity, $\rho$ the fluid density, and $\nu$ be the kinematic viscosity of the fluid surrounding a particle.  The Reynolds number in this scenario is: $\Rey_p = Ua / \nu$.  The drag $F_D$ on the sphere is the force in the $z-$direction.  We define the drag coefficient to be:
\begin{equation*}
  C_D = \frac{F_D}{\rho U^2 a^2}~.
\end{equation*}

In our simulation, we consider the domain of fluid to be a cube of length $50a$, with the particle of radius $a$ centered at the origin.  We choose the minimum element size at the sphere surface to be comparable to those of the simulations described in \S\ref{sec:eqns}. 

We compute the drag force using the Lagrange multipliers used within the FEM to enforce the velocity boundary condition on the particle surface.  We consider Reynolds numbers between $\Rey_p = 0.1$ and $\Rey_p = 100$, by varying the fluid velocity $U$. Our computation of the drag coefficient $C_D$ compares favorably to those of various experimental and numerical studies (figure \ref{fig:dragvRe}$b$). In particular: \cite{Maxworthy65} accurately measures the drag on a sphere in experiments, using a container diameter which is 700 times the sphere diameter.  Maxworthy estimates his experimental error to be better than 2\%.  We also include experimental data catalogued in \cite{perry1950} for larger Reynolds numbers and  numerical studies by \cite{LeClair72} and \cite{Dennis71}.

\section{Analytic velocities}\label{appVel}

This appendix contains the full equations for the velocities, $\bu^{(0)}_1$ and $\mathbf{\hat{u}}_1$, described in \S \ref{sec:asymptotics}.  Since the odd terms of the expansion of $\bu^{(0)}$ are exact on the particle, they can be computed analytically using Lamb's solution for the flow external to a sphere \citep{Lamb45,HoLeal74}.  The multipole expansion of $\bu^{(0)}_1$ takes the form:

\begin{equation}
  \bu^{(0)}_1 = \sum_{n=0}^{\infty} \frac{1}{r^{n+1}} \bfn^1\left(\frac{x- x_0}{r}, \frac{y- y_0}{r},\frac{z}{r}\right).
\end{equation}

The components of each $\bfn^1$ are defined as follows: 

\begin{eqnarray}
  \boldsymbol{\mathrm{f}}_0^1\left(\frac{x}{r},\frac{y}{r},\frac{z}{r}\right) &=&  -\frac{(A_1+I_1)}{2}\left(\mathbf{e}_z+\frac{z\mathbf{r}}{r^2}\right), \\
  \boldsymbol{\mathrm{f}}_1^1\left(\frac{x}{r},\frac{y}{r},\frac{z}{r}\right) &=& \frac{-5\gamma_y}{2} \frac{yz\mathbf{r}}{r^3} + C_1\left(\frac{z\mathbf{e}_y}{r}-\frac{y\mathbf{e}_z}{r}\right) + \frac{-5\gamma_x}{2} \frac{xz\mathbf{r}}{r^3} + K_1\left(\frac{z\mathbf{e}_x}{r} - \frac{x\mathbf{e}_z}{r}\right), \\
  \boldsymbol{\mathrm{f}}_2^1\left(\frac{x}{r},\frac{y}{r},\frac{z}{r}\right) &=& -\frac{(\delta_{yy}+\delta_{xx})}{15}\left(\mathbf{e}_z - \frac{3z\mathbf{r}}{r^2}\right) + \frac{(-\delta_{yy}+\delta_{xx})}{3}\left(\frac{z\mathbf{r}}{r^2} -\frac{(r^2-2y^2)\mathbf{e}_z}{r^2} \right) \\
  &+& \frac{7\delta_{yy}}{120}\left[ \left(13 - \frac{75y^2}{r^2}\right) \frac{z\mathbf{r}}{r^2} + \frac{10yz\mathbf{e}_y}{r^2} -\left(1-\frac{5y^2}{r^2}\right)\mathbf{e}_z  \right] \nonumber \\
  &+& \frac{7\delta_{xx}}{120}\left[ \left(13 - \frac{75x^2}{r^2}\right) \frac{z\mathbf{r}}{r^2} + \frac{10xz\mathbf{e}_x}{r^2} -\left(1-\frac{5x^2}{r^2}\right)\mathbf{e}_z  \right], \nonumber \\
  \boldsymbol{\mathrm{f}}_3^1\left(\frac{x}{r},\frac{y}{r},\frac{z}{r}\right) &=& \frac{-\gamma_y}{2}\left(\frac{z\mathbf{e}_y}{r}+\frac{y\mathbf{e}_z}{r} -\frac{5yz\mathbf{r}}{r^3}\right) + \frac{-\gamma_x}{2}\left(\frac{z\mathbf{e}_x}{r}+\frac{x\mathbf{e}_z}{r} -\frac{5xz\mathbf{r}}{r^3}\right),  \\
  \boldsymbol{\mathrm{f}}_4^1\left(\frac{x}{r},\frac{y}{r},\frac{z}{r}\right) &=& \frac{\delta_{yy}}{8}\left[ -5\left( 1-\frac{7y^2}{r^2} \right)\frac{z\mathbf{r}}{r^2} + \frac{2yz\mathbf{e}_y}{r^2} + \left(1-\frac{5y^2}{r^2}\right)\mathbf{e}_z  \right] \\
  &+& \frac{\delta_{xx}}{8}\left[ -5\left( 1-\frac{7x^2}{r^2} \right)\frac{z\mathbf{r}}{r^2} + \frac{2xz\mathbf{e}_x}{r^2} + \left(1-\frac{5x^2}{r^2}\right)\mathbf{e}_z  \right], \nonumber
\end{eqnarray}
and $\bfn^1 = \mathbf{0}$ for $n\ge 5$. Here the constants $A_1$, $C_1$, $I_1$, and $K_1$ are all of order $O(\alpha^3)$, and so do not participate in determining the force on the particle at the order computed in this study. Note that when when we are on the symmetry plane $x=0$, then also $\gamma_x=0$, and likewise with $y$ and $\gamma_y$.

We can also use the Lamb's solution to calculate the odd terms in the expansion of $\hbu$. In particular, we represent $\mathbf{\hat{u}}_1$ in the following multipole expansion:

\begin{equation}
  \hbu_1 = \sum_{n=0}^{\infty} \frac{1}{r^{n+1}} \boldsymbol{\mathrm{g}}_n^1\left(\frac{x-x_0}{r}, \frac{y-y_0}{r},\frac{z}{r}\right).
\end{equation}

The full analytic solutions for the $\boldsymbol{\mathrm{g}}_n^1$ are below:

\begin{eqnarray} 
 \boldsymbol{\mathrm{g}}_0^1\left(\frac{x}{r},\frac{y}{r},\frac{z}{r}\right) &=&  \frac{3}{4}\left(\mathbf{e}_y+\frac{y\mathbf{r}}{r^2}\right), \label{eq:g01} \\
  \boldsymbol{\mathrm{g}}_2^1\left(\frac{x}{r},\frac{y}{r},\frac{z}{r}\right) &=& \frac{1}{4}\left(\mathbf{e}_y - \frac{3y\mathbf{r}}{r^2}\right),
\end{eqnarray}

and $\boldsymbol{\mathrm{g}}_n^1=\mathbf{0}$ for $n=1$, and $n\ge3$.

\section{Determining the reciprocal theorem integrands}\label{appIntegrand}

In this appendix, we rationalize the choice of integrands for the reciprocal theorem (\ref{eq:recipthm}) in \S \ref{sec:innerint} and \S\ref{sec:outerint}.  For each domain, we scale by the characteristic length, and then sort terms by magnitude in $\alpha$.  Finally, we choose terms of the velocities that combine to give the desired power of $\alpha$.

\subsection{Inner integral}

For the inner integral, we continue to scale lengths by the particle radius $a$, and collect terms by order of magnitude in $\alpha$.  For the $O(a^4)$ contribution, we need to choose combinations of $\bu^{(0)}_i$, $\hbu_i$, and $\mathbf{\bar{u}}$ that combine to give $O(\alpha^2)$ in the integrand of (\ref{eq:recipthm}).  Similarly, for $O(a^5)$, terms need to combine to give $O(\alpha^3)$ in the integrand. 

The leading terms in magnitude $\alpha$ of the $\bu^{(0)}_i$ are shown below.

\begin{equation}
  \begin{aligned}
    \bu^{(0)}_1 &\sim \alpha \left(\frac{1}{r^2}\boldsymbol{\mathrm{f}}_1^1 + \frac{1}{r^4}\boldsymbol{\mathrm{f}}_3^1\right) + \alpha^2 \left(\frac{1}{r^3}\boldsymbol{\mathrm{f}}_2^1  + \frac{1}{r^5}\boldsymbol{\mathrm{f}}_4^1\right) +O(\alpha^4), \\
    \bu^{(0)}_2 &\sim \alpha^3 \mathcal{SI}\left[\frac{1}{r^2}\boldsymbol{\mathrm{f}}_1^1\right] + O(\alpha^4),
  \end{aligned}
\end{equation}
and all higher order $\bu^{(0)}_i$ are $O(\alpha^4)$ or smaller. We define $\mathbf{v}=\mathcal{SI}[\bu]$ as the image of the function $\bu$, that is $\mathbf{v}$ solves the Stokes equations with $-\bu$ as the boundary condition on the walls.  The leading terms in magnitude $\alpha$ of the $\hbu_i$ are shown below.

\begin{equation}
  \begin{aligned}
    \hbu_1 &\sim \frac{1}{r}\boldsymbol{\mathrm{g}}_0^1 + \frac{1}{r^3}\boldsymbol{\mathrm{g}}_2^1 ~,\\
    \hbu_2 &\sim \alpha \mathcal{SI}\left[ \frac{1}{r}\boldsymbol{\mathrm{g}}_0^1 \right] + O(\alpha^3)~, \\
    \hbu_3 &\sim \alpha \left( \frac{1}{r}\boldsymbol{\mathrm{g}}_0^3+ \frac{1}{r^3}\boldsymbol{\mathrm{g}}_2^3 \right) ,
  \end{aligned}
\end{equation}
and higher order $\hbu_i$ are $O(\alpha^2)$ or smaller. The leading terms in magnitude $\alpha$ of $\mathbf{\bar{u}}$ are:
\begin{equation}
  \mathbf{\bar{u}} \sim [\alpha \gamma r + \alpha^2 \delta r^2 + O(\alpha)]\mathbf{e_z}~.
\end{equation}

Recall that the inner integral has the $\alpha$ expansion:
\begin{equation}
  f_{L_1} = \rho U_m^2 a^2 ( h_4\alpha^2 + h_5\alpha^5 + \hdots \,)~.
\end{equation}
It is evident that only $\bu^{(0)}_1$, $\hbu_1$, and the shear term of $\mathbf{\bar{u}}$ (call it $\bar{\bu}_{\gamma}$) contribute to the $O(a^4)$ term of the inner integral, that is:

\begin{equation}
  h_4 = \int_{\mathbb{R}^3} \hbu_1 \cdot (\bar{\bu}_{\gamma} \cdot \nabla \bu^{(0)}_1 + \bu^{(0)}_1 \cdot \nabla \bar{\bu}_{\gamma}) \,\mathrm{dv}.
\end{equation}
Whereas $\bu^{(0)}_1$, $\hbu_1$, $\hbu_2$, $\hbu_3$, and both the shear and curvature terms of $\mathbf{\bar{u}}$ (call them $\bar{\bu}_{\gamma}$ and $\bar{\bu}_{\delta}$ respectively), contribute to the $O(a^5)$ term.

\begin{equation}
  h_5 = \int_{\mathbb{R}^3} (\hbu_1 + \hbu_2 + \hbu_3) \cdot \left[(\bar{\bu}_{\gamma} + \bar{\bu}_{\delta}) \cdot \nabla \bu^{(0)}_1 + \bu^{(0)}_1 \cdot \nabla (\bar{\bu}_{\gamma} + \bar{\bu}_{\delta})\right] \,\mathrm{dv}.
\end{equation}
These integrals are evaluated in \S \ref{sec:innerint}.

\subsection{Outer integral}

The outer integral is expressed in terms of $\mathbf{R}=\alpha\mathbf{r}$.  We arrange our functions in order of magnitude in $\alpha$ as shown below.  For the $O(a^4)$ term in the outer integral, we need to collect terms of $\mathbf{U}^{(0)}_i$, $\mathbf{\hat{U}}_i$, and $\mathbf{\bar{U}}$ which combine to give $O(\alpha^4)$ in the integrand of (\ref{eq:recipthm}).  For the $O(a^5)$ term, we need an $O(\alpha^5)$ integrand in (\ref{eq:recipthm}).

The leading terms in magnitude $\alpha$ of the $\mathbf{U}^{(0)}_i$ are:

\begin{equation}
\begin{aligned}
  \mathbf{U}^{(0)}_1 &\sim \alpha^3 \left( \frac{1}{R^2} \boldsymbol{\mathrm{f}}_1^1  + \frac{1}{R^4} \boldsymbol{\mathrm{f}}_3^1    \right) + O(\alpha^4)~,\\
  \mathbf{U}^{(0)}_2 &\sim   \alpha^3 \mathcal{SI}\left[\frac{1}{R^2}\boldsymbol{\mathrm{f}}_1^1 \right]   + O(\alpha^4)~,
  \end{aligned}
	\end{equation}
and $\mathbf{U}^{(0)}_3$ and $\mathbf{U}^{(0)}_4$ are $O(\alpha^4)$.  The leading terms in magnitude $\alpha$ of the $\mathbf{\hat{U}}_i$ are:
  
\begin{equation}
\begin{aligned}
  \mathbf{\hat{U}}_1 &\sim \alpha \frac{1}{R} \boldsymbol{\mathrm{g}}_0^1   +  \alpha^3 \frac{1}{R^3} \boldsymbol{\mathrm{g}}_2^1~,   \\
  \mathbf{\hat{U}}_2 &\sim \alpha \mathcal{SI}\left[\frac{1}{R}\boldsymbol{\mathrm{g}}_0^1 \right]  + O(\alpha^3)~, \\
  \mathbf{\hat{U}}_3 &\sim  \alpha^2 \frac{1}{R} \boldsymbol{\mathrm{g}}_0^3   +  \alpha^4 \frac{1}{R^3} \boldsymbol{\mathrm{g}}_2^3   + O(\alpha^4)~,\\
  \mathbf{\hat{U}}_4 &\sim \alpha^2 \mathcal{SI}\left[\frac{1}{R}\boldsymbol{\mathrm{g}}_0^3 \right] + O(\alpha^4)~,
\end{aligned}
\end{equation}
while all the terms of $\mathbf{\bar{U}}$ are $O(1)$.  Recall that the outer integral has the $\alpha$ expansion:
\begin{equation}
  f_{L_2} = \rho U_m^2 \ell^2 ( k_4 \alpha^4 + k_5 \alpha^5 + \hdots \,)~.
\end{equation}
Only $\mathbf{U}^{(0)}_1$, $\mathbf{U}^{(0)}_2$, $\mathbf{\hat{U}}_1$, $\mathbf{\hat{U}}_2$, and $\mathbf{\bar{U}}$ contribute to the $O(a^4)$ term in the outer integral, namely:

\begin{equation}
  k_4 = \int_{V_C} (\mathbf{\hat{U}}_1 + \mathbf{\hat{U}}_2) \cdot \left[ \bar{\mathbf{U}} \cdot \nabla (\mathbf{U}^{(0)}_1 + \mathbf{U}^{(0)}_2) + (\mathbf{U}^{(0)}_1 + \mathbf{U}^{(0)}_2) \cdot \nabla \bar{\mathbf{U}}  \right] \,\mathrm{dv}.
\end{equation}
Whereas $\mathbf{U}^{(0)}_1$, $\mathbf{U}^{(0)}_2$, $\mathbf{\hat{U}}_3$, $\mathbf{\hat{U}}_4$, and $\mathbf{\bar{U}}$ contribute to the $O(a^5)$ term:

\begin{equation}
  k_5 = \int_{V_C} (\mathbf{\hat{U}}_3 + \mathbf{\hat{U}}_4) \cdot \left[ \bar{\mathbf{U}} \cdot \nabla (\mathbf{U}^{(0)}_1 + \mathbf{U}^{(0)}_2) + (\mathbf{U}^{(0)}_1 + \mathbf{U}^{(0)}_2) \cdot \nabla \bar{\mathbf{U}}  \right] \,\mathrm{dv}.
\end{equation}
These integrals are evaluated in \S \ref{sec:outerint}.

\bibliographystyle{jfm}

\bibliography{inertial-migration}

\end{document}